\documentclass[10pt, conference, compsocconf]{IEEEtran}
\usepackage{setspace}
\usepackage{multirow}
\usepackage{graphicx}
\usepackage{rotating}
\usepackage{float}
\usepackage{verbatim}
\usepackage{courier}
\usepackage{flushend}
\usepackage{cite}
\usepackage{amsmath,amssymb,amsfonts}
\usepackage{algorithm}
\usepackage{textcomp}
\usepackage{xcolor}
\usepackage{balance}
\usepackage{listings}
\usepackage[belowskip=-12pt,aboveskip=10.5pt]{caption}
\usepackage{wrapfig}

\usepackage{algpseudocode}  

\usepackage{subcaption}

\colorlet{punct}{red!60!black}
\definecolor{background}{HTML}{EEEEEE}
\definecolor{delim}{RGB}{20,105,176}
\colorlet{numb}{magenta!60!black}
\lstdefinelanguage{json}{
    basicstyle=\normalfont\ttfamily\small,
    numbers=left,
    numberstyle=\scriptsize,
    stepnumber=1,
    numbersep=8pt,
    showstringspaces=false,
    breaklines=true,
    frame=lines,
    backgroundcolor=\color{background},
    literate=
     *{0}{{{\color{numb}0}}}{1}
      {1}{{{\color{numb}1}}}{1}
      {2}{{{\color{numb}2}}}{1}
      {3}{{{\color{numb}3}}}{1}
      {4}{{{\color{numb}4}}}{1}
      {5}{{{\color{numb}5}}}{1}
      {6}{{{\color{numb}6}}}{1}
      {7}{{{\color{numb}7}}}{1}
      {8}{{{\color{numb}8}}}{1}
      {9}{{{\color{numb}9}}}{1}
      {:}{{{\color{punct}{:}}}}{1}
      {,}{{{\color{punct}{,}}}}{1}
      {\{}{{{\color{delim}{\{}}}}{1}
      {\}}{{{\color{delim}{\}}}}}{1}
      {[}{{{\color{delim}{[}}}}{1}
      {]}{{{\color{delim}{]}}}}{1},
}

\ifCLASSINFOpdf
\else
\fi
\usepackage{float}
\usepackage{url}

\usepackage{pifont}

\newcommand{\smallcapital}{\fontsize{9pt}{10pt}\selectfont}
\begin{document}
%
\title{$\mu$qSim: Scalable and Validated Simulation of Cloud Microservices}


\author{\IEEEauthorblockN{Authors Name/s per 1st Affiliation (Author)}
\IEEEauthorblockA{line 1 (of Affiliation): dept. name of organization\\
line 2: name of organization, acronyms acceptable\\
line 3: City, Country\\
line 4: Email: name@xyz.com}
\and
\IEEEauthorblockN{Authors Name/s per 2nd Affiliation (Author)}
\IEEEauthorblockA{line 1 (of Affiliation): dept. name of organization\\
line 2: name of organization, acronyms acceptable\\
line 3: City, Country\\
line 4: Email: name@xyz.com}
}

\author{\IEEEauthorblockN{Yanqi Zhang, Yu Gan, and Christina Delimitrou}
\IEEEauthorblockA{Electrical and Computer Engineering Department\\
Cornell University\\
\{yz2297, yg397, delimitrou\}@cornell.edu}}


%


\maketitle
\thispagestyle{plain}
\pagestyle{plain}

\begin{abstract}
{ Current cloud services are moving away from monolithic designs and towards graphs 
of many loosely-coupled, single-concerned microservices. Microservices have several advantages, 
including speeding up development and deployment, allowing specialization of the software infrastructure, 
and helping with debugging and error isolation. At the same time they introduce several hardware and software challenges. 
Given that most of the performance and efficiency implications of microservices happen at scales larger than what is available 
outside production deployments, studying such effects requires designing the right simulation infrastructures. 

We present {\bf$\mu$}qSim, a scalable and validated queueing network simulator designed specifically for interactive microservices. 
$\mu$qSim provides detailed intra- and inter-microservice models that allow it to faithfully reproduce the behavior of complex, many-tier 
applications. $\mu$qSim is also modular, allowing reuse of individual models across microservices and end-to-end applications. 
We have validated $\mu$qSim both against simple and more complex microservices graphs, and have shown that it accurately captures performance 
in terms of throughput and tail latency. Finally, we use $\mu$qSim to model the tail at scale effects of request fanout, and the performance impact 
of power management in latency-sensitive microservices. 
 }

\end{abstract}


%
\IEEEpeerreviewmaketitle

\section{Introduction}

An increasing amount of computing is now performed in the cloud, primarily
due to the resource flexibility benefits for end users, and the cost benefits for both end users and cloud providers~\cite{BarrosoBook,Barroso11,GoogleTrace}. 
Users obtain resource flexibility by scaling their resources on demand and being charged only for the time these resources are used, and 
cloud operators achieve cost efficiency by multiplexing their infrastructure across users~\cite{Lo15,Delimitrou13,Delimitrou14,Delimitrou14b,Delimitrou16,Delimitrou17,Delimitrou13e,Mars13b,Nathuji10,Nathuji07,Mars11a,Mars13a}. 

Most of the services hosted in datacenters are governed by strict quality of service (QoS) 
constraints in terms of throughput and tail latency, as well as availability 
and reliability guarantees~\cite{tailatscale,Delimitrou13,Delimitrou13d,Delimitrou14}. 
In order to satisfy these often conflicting requirements, cloud services have seen a significant shift in their design, moving away 
from the traditional monolithic applications, where a single service encompasses 
the entire functionality, and adopting instead a multi-tier microservices application model~\cite{Cockroft15,Cockroft16}. 
Microservices correspond to fine-grained, single-concerned and loosely-coupled application tiers, 
which assemble to implement more sophisticated functionality~\cite{Cockroft15,twitter_decomposing,Ueda16,Sriraman18,Zhou18,Ppbench,Gan19,Gan18,Gan18b,Delimitrou19}. 

Microservices are appealing for several reasons. First, they improve programmability by simplifying and accelerating deployment through modularity. 
Unlike monolithic services, each microservice is responsible for a small, well-defined
fraction of the entire application's functionality, with different microservices being independently deployed. 
Second, microservices can take advantage of language and programming framework heterogeneity, since they only
require a common cross-application API, typically over remote procedure calls (RPC) or a RESTful API~\cite{thrift}.
Third, individual microservices can easily be updated, or swapped out and replaced by newer modules without major changes to the rest of the application's architecture. 
In contrast, \textit{monoliths} make frequent updates cumbersome and error-prone, and limit the set of programming languages that can be used for development.

\begin{figure}
\vspace{-0.08in}
\centering
\includegraphics[scale=0.344]{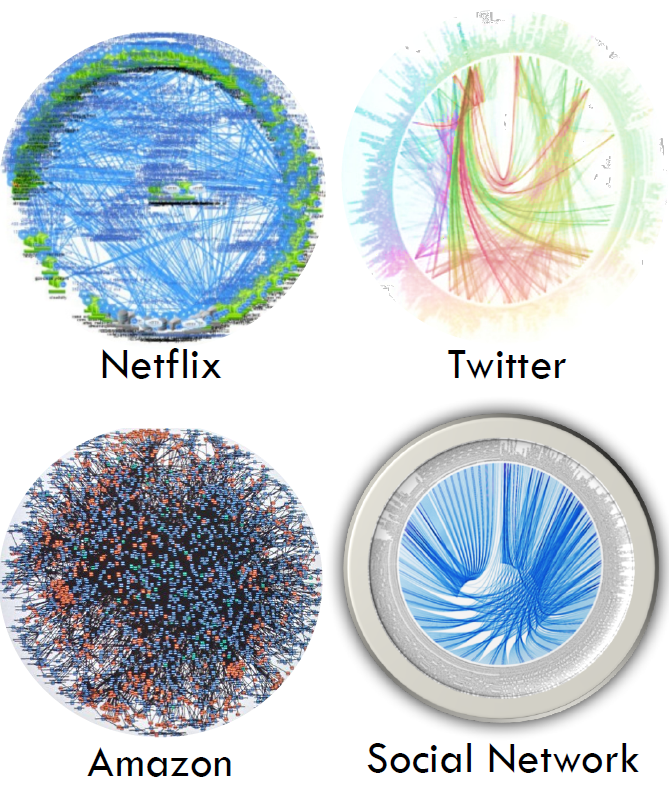}
\caption{{\bf{\label{fig:microservicesGraphs}}} {Microservices graphs in three large
cloud providers~\cite{Cockroft15,Cockroft16,twitter_decomposing}, and our \texttt{Social Network} service. }}
\end{figure}

Fourth, microservices simplify correctness and performance debugging, as bugs can be isolated
to specific components, unlike monoliths, where troubleshooting often involves the end-to-end service.
Finally, microservices fit nicely the model of a container-based datacenter, with a microservice per container, and microservices being 
scaled independently according to their resource demands. 
An increasing number of cloud service providers, including Twitter, Netflix, AT\&T, Amazon, and eBay have adopted this application model~\cite{Cockroft15}.

Despite their advantages, microservices introduce several hardware and software challenges. 
First, they put increased pressure on the network system, as dependent microservices reside in different physical nodes, relying on an RPC or HTTP layer for communication. 
Second, microservices further complicate cluster management, as instead of allocating resources to a single service, the scheduler must now determine the impact of dependencies 
between any two microservices in order to guarantee end-to-end QoS. Microservices further exacerbate tail-at-scale effects~\cite{tailatscale} as a single poorly-configured 
microservices on the critical path can cause cascading QoS violations resulting in end-to-end performance degradation. 
Typical dependency graphs between microservices in production systems involve several hundred microservices; 
Fig.~\ref{fig:microservicesGraphs} shows a few representative examples from Netflix, Twitter, Amazon~\cite{Cockroft15}, and one of the applications used later in this paper.

Given the fact that unpredictable performance often only emerges at large scale, it is critical to study the resource management, availability, and responsiveness of 
microservices at scales larger than what is possible in typical research facilities. 
Towards this effort we leverage the following insight: a positive side-effect of the simplicity of individual microservices is that - unlike complex monoliths - they conform 
to the principles of queueing theory. In this work we use this insight 
to enable accurate microservices simulation that relies on queueing networks to capture the impact of dependencies between microservices.  

We present $\mu$qSim, an accurate and scalable simulator for interactive cloud microservices. $\mu$qSim captures dependencies between individual microservices, as well as 
application semantics such as request batching, blocking connections, and request pipelining. $\mu$qSim relies on a user-provided high-level declarative specification of 
the microservices dependency graph, and a description of the available server platforms. $\mu$qSim is an event-driven simulator, and uses a centralized scheduler 
to dispatch requests to the appropriate microservices instances. 

We validate $\mu$qSim against both simple and complex microservices graphs, including multi-tier web applications, services with load balancing, and services with high request fanout, 
where all leaf nodes must respond before the result is returned to the user. We show that $\mu$qSim faithfully reproduces the performance (throughput and tail latency) of 
the real application in all cases. 
We then use $\mu$qSim to evaluate two use cases for interactive microservices. First, we show that tail at scale effects become worse in microservices compared to single-tier services, 
as a single under-performing microservice on the critical path can degrade end-to-end performance. 
Second, we design and evaluate a QoS-aware power management mechanism for microservices both in a real server and in simulation. The power manager 
determines the appropriate per-microservice QoS requirements needed to meet the end-to-end performance constraints, and dynamically adjusts frequency accordingly. 




\section{Related Work}
\label{sec:related_work}

Queueuing network simulators are often used to gain insight on performance/resource trade-offs in systems that 
care about request latency \cite{meisner2012bighouse,zeng1998glomosim,xie2009aqua,henderson2008network,issariyakul2012introduction}. The closest system to $\mu$qSim is BigHouse, an event-driven queueing simulator
targeting datacenter service simulation. 
BigHouse represents workloads as inter-arrival and service distributions, whose characteristics are obtained through offline profiling and online 
instrumentation. The simulator then models each application as a single queue, 
and runs multiple instances in parallel until performance metrics converge. 
While this offers high simulation speed, and can be accurate for simple applications, it introduces non-negligible errors 
when simulating microservices because intra-microservice stages cannot be captured by a single queue. For example, a request to an application like memcached 
must traverse multiple stages before completion, including network processing, e.g., TCP/IP rx, 
processing in OS event-driven libraries like \texttt{epoll} and \texttt{libevent}, reading the received requests from the socket, and finally 
processing the user request. Along each of these stages there are queues, and ignoring them can underestimate 
the impact of queueing on tail latency. 
Furthermore, BigHouse is only able to model single tier applications. Microservices typically consist of multiple tiers, and 
inter-tier dependencies are not a straightforward pipeline, often involving blocking, synchronous, or cyclic communication between microservices.

$\mu$qSim takes a different approach by explicitly modeling each application's execution stages, 
and accounting for queueing effects throughout execution, including request batching, e.g., in \texttt{epoll}, 
request blocking, e.g., in http 1/1.1 \& disk I/O, and request parallelism, e.g., across threads or processes. 
Given that many of these sources of queueing are repeated across microservices, $\mu$qSim provides 
modular models of application stages, which can be reused across diverse microservices. 
Moreover, $\mu$qSim supports user-defined microservice graph topology and flexible blocking and synchronization behaviors between tiers of microservices.


\begin{figure}
\centering
\includegraphics[scale=0.40,bb=100 60 500 450]{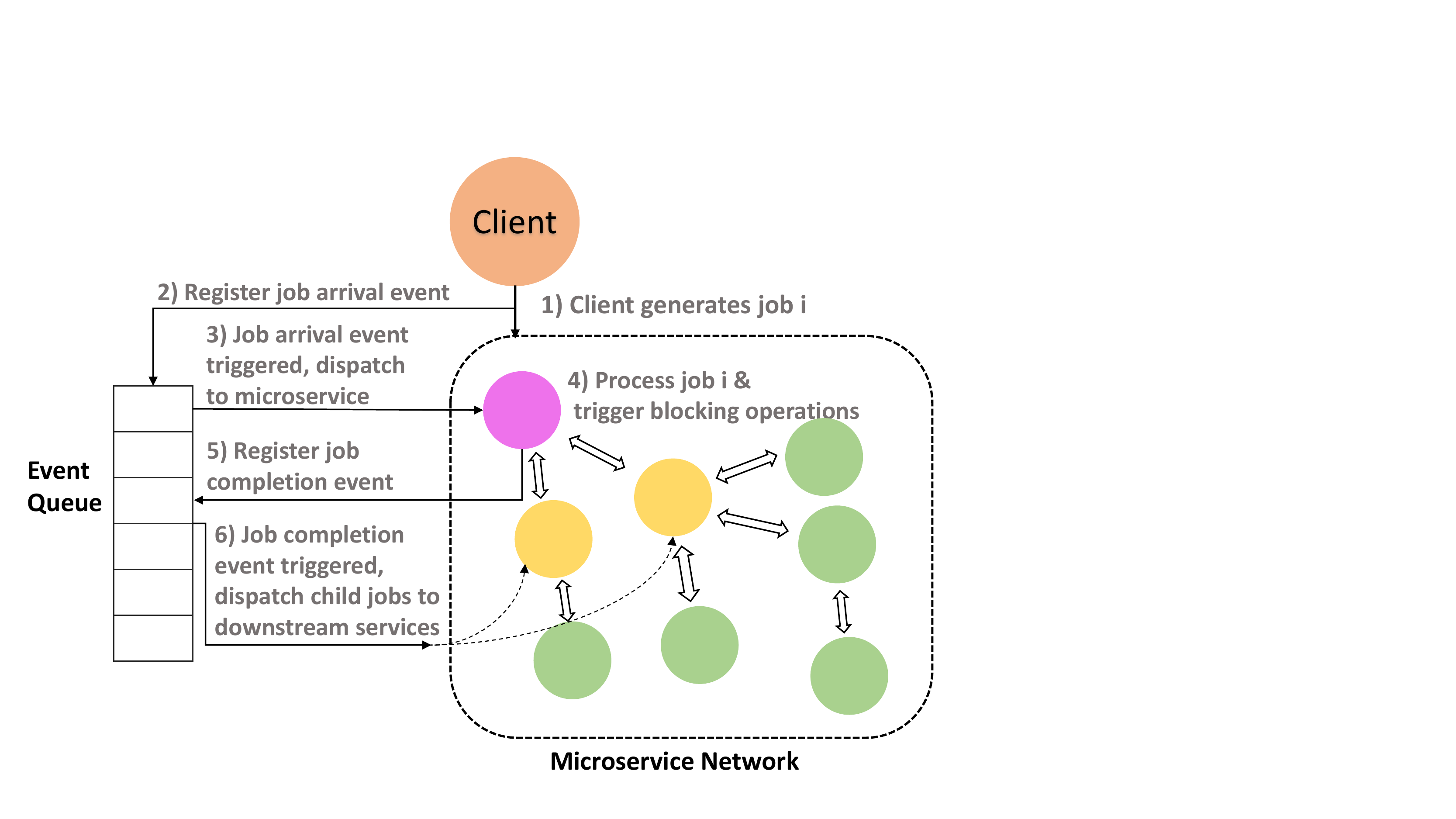}
\caption{\label{fig:simulator} Overview of the event-driven behavior in $\mu$qSim. }
\end{figure}

\section{Design}

$\mu$qSim is designed to enable modularity and reuse of models across microservices and end-to-end applications, as well as to allow enough 
flexibility for users to incorporate their own applications in the simulator. 
The core of $\mu$qSim is implemented in standard C++ with 25,000 lines of codes. 
The simulator interface requires the user to express a microservice's internal logic 
using a {\smallcapital JSON} configuration file, similar to the 
one shown in~\ref{lst:memc_json}. The user additionally needs to express the end-to-end service's 
architecture, i.e., how are different microservices connected to each other, 
and what hardware resources are available in the cluster. 
We summarize the input users express to the simulator in Table~\ref{tab:user_input}.
Below we detail each of these components. 

\begin{table}[]
\small
\centering
\caption{Simulation inputs. }
\label{tab:user_input}
\begin{tabular}{l|l}
\hline
\hline
\texttt{service.json}                       & Internal architecture of a microservice     \\
\texttt{graph.json}                         & Inter-microservice topology \\
\texttt{path.json}     & Paths (sequence of microservices) \\
		       & that requests follow across microservices \\
\texttt{machines.json}                      & Server machines \& available resources                                    \\
\texttt{client.json}                        & Input load pattern                                              \\
\texttt{histograms}                    & Processing time PDF per microservice                                    \\
\hline
\hline
\end{tabular}
\end{table}


\subsection{Simulation Engine}

$\mu$qSim uses discrete event simulation, as shown in Fig.~\ref{fig:simulator}. An event may represent the arrival or completion of a job 
(a request in a microservice), as well as cluster administration operations, like changing a server's {\smallcapital DVFS} setting. 
Each event is associated with a timestamp, and all events are stored in increasing time order in a priority queue. In every simulation cycle, 
the simulation queue manager queries the priority queue for the earliest event. It then uses the microservice model the event corresponds to 
to compute the execution time and resources required to process the event, as well as to insert causally dependent events to the priority queue. 
These include requests triggered in other microservices as a result of the processed event. 
Simulation completes when there are no more outstanding events in the priority queue.

\subsection{Modeling Individual Microservices}

$\mu$qSim models each individual microservice with two orthogonal components: \textit{application logic} and \textit{execution model}. 
The application logic captures the behavior of a microservice. The basic element of the application logic is a stage, 
which represents an execution phase within the microservice, and is essentially a queue-consumer pair, as defined in queueing theory~\cite{Harchol13,Kleinrock}. 
Each stage can be configured with different queueing features like batching or pipelining, and is coupled with a job queue. For example,
the \texttt{epoll} stage is coupled with multiple subqueues, one per network connection.
A stage is also assigned to one or more execution time distributions that describe the stage's processing time under different settings, 
like different {\smallcapital DVFS} configurations, different loads and different thread counts. $\mu$qSim supports processing
time expressed using regular distributions, such as exponential, or via processing time histograms collected through profiling, which requires  
users to instrument applications and record timestamps at boundaries of queueing stages. 

Multiple application logic stages are assembled to form execution paths, corresponding to a microservice's different code paths. 
Finally, the model of a microservice also includes a state machine that specifies the probability that a microservice follows different execution paths. 




\begin{figure}
\centering
\includegraphics[scale=0.35,bb=200 100 600 550]{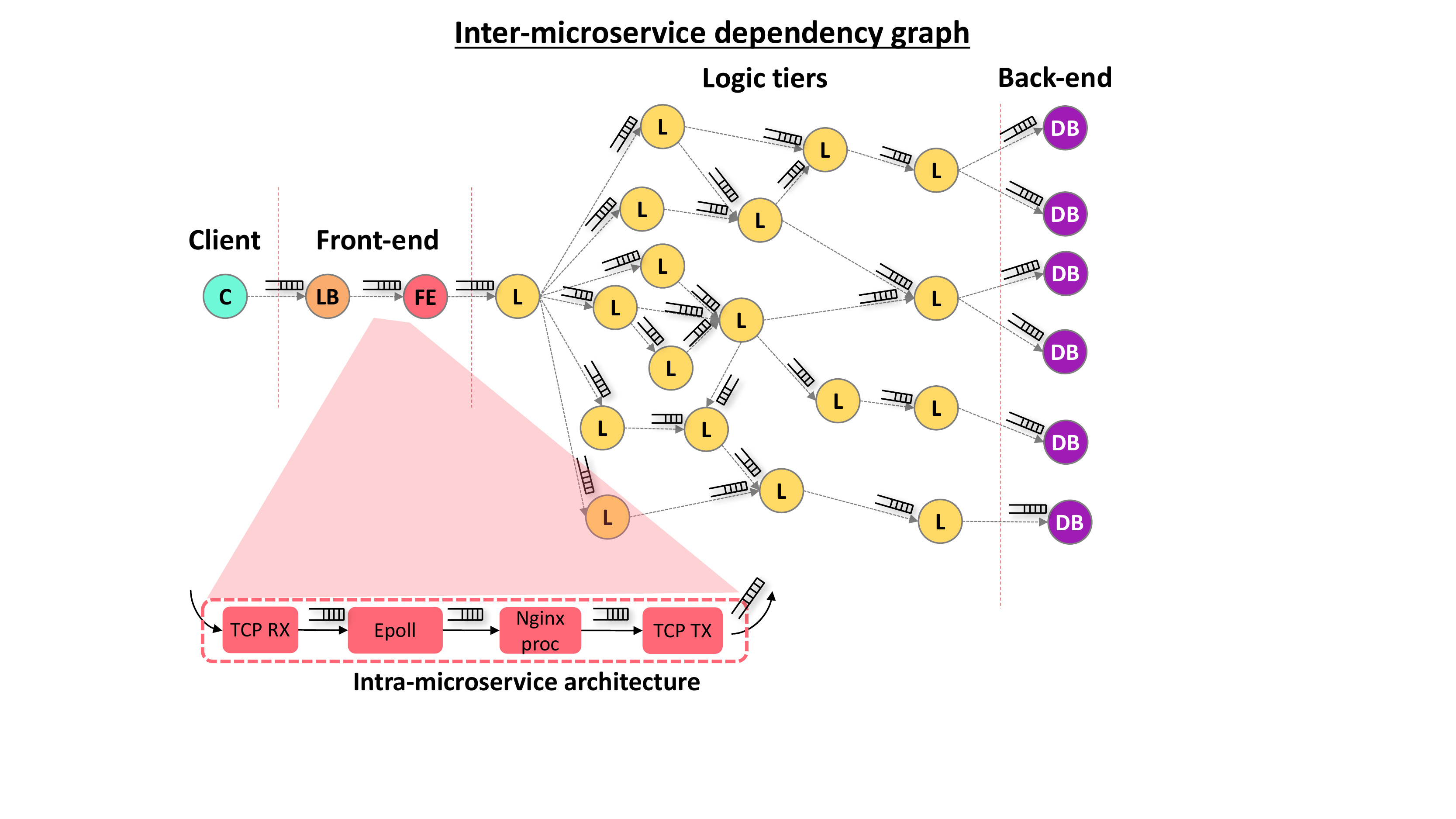}
\caption{\label{fig:inter_overview} Overview of the modeling of an end-to-end service built with microservices in $\mu$qSim (top), and modeling of the execution stages within a single microservice (bottom). {\smallcapital C}: client, {\smallcapital LB}: load balancer, 
{\smallcapital FE}: front-end, {\smallcapital L}: logic tier, {\smallcapital DB}: database tier. }
\end{figure}

Fig.~\ref{lst:memc_json} shows memcached's application logic and execution path using $\mu$qSim's {\smallcapital JSON} template. 
The main stages include \texttt{epoll}, \texttt{socket\_read}, and \texttt{memcached\_processing}. This excludes network processing, which 
is modeled as a separate process in the simulator: each server is coupled with a network processing process as a standalone service, 
and all microservices deployed on the same server share the processed handling interrupts. 
Both \texttt{epoll} and \texttt{socket\_read} use request batching, and can return more than one jobs. 
The \texttt{epoll} queue classifies jobs into subqueues based on socket connections, 
and returns the first $N$ jobs of each active subqueue (defined in ``queue\_parameter''). 
\texttt{socket\_read} similarly classifies jobs based on connections, 
but returns the first $N$ jobs from a single ready connection at a time. 
The \texttt{memcached\_processing} and \texttt{socket\_send} stages do not use batching, and their queues simply store all jobs in one queue. 

It is important to note that the processing time of \texttt{epoll} and \texttt{socket\_read} are runtime dependent: 
\texttt{epoll}'s execution time increases linearly with the number of active events that are returned, and \texttt{socket\_read}{'s 
processing time is also proportional to the number of bytes read from socket. 
Finally, memcached's execution model consists of two paths, one for \texttt{read} requests, and one for \texttt{write}. 
These two paths consist of exactly the same stages in the same order, and are only used to distinguish between different 
processing time distributions. Each path for memcached is deterministic, therefore there is no need for a probability distribution to select 
an execution path. 
One example where probabilistically selecting execution paths is needed is MongoDB, where a query could be either a cache hit that only accesses memory, 
or a cache miss that results in disk I/O. The probability for each path in that case is a function of MongoDB's working set size and allocated memory.
Currently $\mu$qSim only models applications running on standard Linux. We defer modeling acceleration techniques like user level networking, such as {\smallcapital DPDK} and 
{\smallcapital FPGA} accelerated networking to future work.

The execution model of a microservice also describes how a job is processed by the simulator. 
Currently $\mu$qSim supports two models: simple and multi-threaded. A simple model directly 
dispatches jobs onto hardware resources like {\smallcapital CPU}, and is mainly used for simple (single stage) services. 
Multi-threaded models add the abstraction of a thread or process, which users can specify either statically or using a 
dynamic thread/process spawning policy. In a multi-threaded model, a job will be first dispatched to a thread, 
and the microservice will search for adequate resources to execute the job, or stall if no resources are available. 
The multi-threaded model captures context switching and I/O blocking overheads, 
and is typically used for microservices with multiple execution stages that include blocking/synchronization.

\begin{table}[]
	\small
	\centering
	\caption{Specification of the server used for all validation experiments of $\mu$qSim. }
	\label{tab:server_spec}
	  \begin{tabular}{ll}
		            \hline
			              \hline
				          Model                  & Intel(R) Xeon(R) CPU E5-2660 v3                 \\
				          OS                     & Ubuntu 14.04(kernel 4.4.0) \\
				          Sockets                & 2                                                \\
				          Cores/Socket           & 10                                               \\
				          Threads/Core           & 2                                                \\
				          Min/Max DVFS Freq.	 & 1.2GHz/2.6GHz                                    \\
				          L1 Inst/Data Cache     & 32KB/32KB                                        \\
				          L2 Cache               & 256KB                                            \\
				          L3 (Last-Level) Cache  & 25.6MB                                           \\
				          Memory                 & 8x16GB 2400MHz DDR4                              \\
				          Hard Drives             & 2x 2TB 7.2K RPM SATA                              \\
				          Network Bandwidth      & 1Gbps       \\
				                \hline
						          \hline
							    \end{tabular}
						    \end{table}

\subsection{Modeling a Microservice Architecture}

A microservice architecture is specified in three {\smallcapital JSON} files that describe the cluster configuration, 
microservice deployment, and inter-microservice logic. The cluster configuration file (\texttt{machines.json}) records the available 
resources on each server. The microservice deployment file (\texttt{graph.json}) specifies the server on which a microservice is deployed -- if specified, 
the resources assigned to each microservice, and the execution model (simple or multi-threaded) each microservice 
is simulated with. The microservice deployment also specifies the size of the connection pool of each microservice, if applicable. 
\begin{lstlisting}[caption={JSON Specification for memcached},label={lst:memc_json},language=json,numbers=none,firstnumber=1,basicstyle=\footnotesize]
{"service_name": "memcached", 
  "stages": [{
      "stage_name":"epoll", "stage_id":0, 
      "queue_type":"epoll", "batching":true, 
      "queue_parameter":[null, N]}, {
      "stage_name":"socket_read", "stage_id":1, 
      "queue_type":"socket", "batching":true, 
      "queue_parameter":[N] }, {
      "stage_name":"memcached_processing", 
      "stage_id":2, "queue_type":"single", 
      "batching":false, "queue_parameter":null}, {
      "stage_name":"socket_send", "stage_id":3, 
      "queue_type":"single", "batching":false, 
      "queue_parameter":null}], 
  "paths": [{
      "path_id":0, "path_name":"memcached_read", 
      "stages":[0, 1, 2, 3]}, {
      "path_id":1, "path_name":"memcached_write", 
      "stages":[0, 1, 2, 3]}]}
\end{lstlisting}

Finally, the inter-microservice path file (\texttt{path.json}) specifies the sequence of individual microservices 
each job needs to go through. Fig.~\ref{fig:inter_overview} shows such a dependency graph between microservices, 
from a client (C), to a load balancer (LB), front-end (FE), logic tiers (L), and back-end databases (DB), 
The same figure also shows the intra-microservice execution path for one of the microservices, 
\texttt{front-end}, implemented in this example using {\smallcapital NGINX}~\cite{nginx}. 
If the application exhibits control flow variability, users can also specify multiple inter-microservice paths, 
and the corresponding probability distribution for them. 
The basic elements of an inter-microservice path are path nodes, 
which are connected in a tree structure and serve three roles: 

\begin{itemize}
	\item \textit{Specify the microservice, the execution path within the microservice, 
		and the order of traversing individual microservices}. When entering a 
		new path node, the job is sent to a microservice 
		instance designated by the path node and connected with the microservices that 
		the job has already traversed, as defined in \texttt{graph.json}. 
		Each path node can have multiple children, and after execution on the 
		current path node is complete, $\mu$qSim makes a copy of the job 
		for each child node, and sends it to a matching microservice instance.

	\item \textit{Express synchronization}. Synchronization primitives are prevalent 
		in microservice architectures, such as in the case where a microservice can 
		start executing a job if and only if it has received all information 
		from its upstream microservices. In $\mu$qSim, synchronization requirements 
		are expressed in terms of the fan-in of each inter-microservice path node: 
		before entering a new path node, a job must wait until execution in 
		all parent nodes is complete. 
		For example, if NGINX\_0 serves as a proxy and NGINX\_1 and NGINX\_2 operate as file servers, 
		for each user request, NGINX\_0 sends requests to both NGINX\_1 and NGINX\_2, 
		waits to synchronize their responses, and then sends the final response to the client. 

	\item \textit{Encode blocking behavior}. Blocking behavior between microservices is 
		common in {\smallcapital RPC} frameworks, http1/1.1 protocols, and I/O accessing. 
		To represent arbitrary blocking behavior, each path node has 
		two operation fields, one upon entering the node and another upon leaving the node, 
		to trigger blocking or unblocking events on a specific connection or thread. 
		Assume for example a two-tier application, with {\smallcapital NGINX} as the 
		front-end webserver, and \texttt{memcached} as the in-memory caching tier. 
		The client queries the webserver over http 1.1. Once a job starts executing, 
		it blocks the receiving side of the incoming connection 
		(since only one outstanding request is allowed per connection in http 1.1). 
		The condition to unblock this path once request processing is complete 
		is also specified in the same {\smallcapital JSON} file. 
		When the job later returns the \texttt{<key,value>} pair from memcached to {\smallcapital NGINX}, 
		$\mu$qSim searches the list of job ids for the one matching 
		the request that initiated the blocking behavior, in order 
		to unblock the connection upon completion of the current request. 
		Users can also specify other blocking primitives, like thread blocking, 
		in $\mu$qSim. 

\end{itemize}

\section{Validation}

\begin{figure}
	\begin{tabular}{cc}
		\includegraphics[scale=0.33,bb=-100 0 200 300]{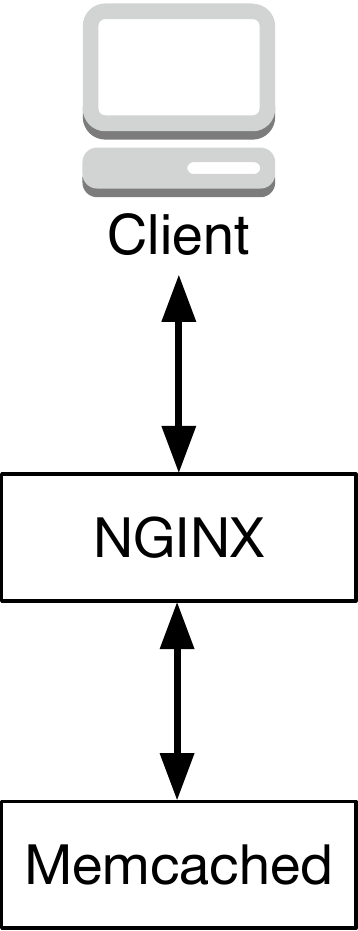} & 
	\includegraphics[scale=0.33,bb=0 0 300 300]{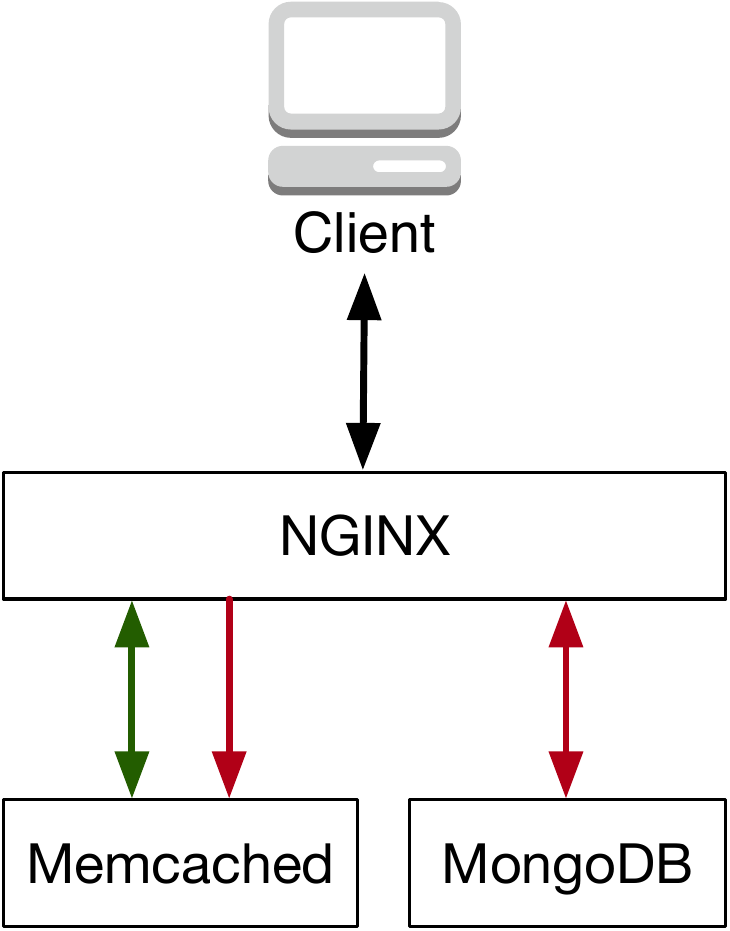}
\end{tabular}
	\caption{The architecture of the 2- and 3-tier applications ({\smallcapital NGINX}-memcached and {\smallcapital NGINX}-memcached-MongoDB). }
  \label{fig:2tier_app}
\end{figure}

\begin{figure*}
  \centering
  \begin{tabular}{cc}
	  \includegraphics[width=0.5\linewidth]{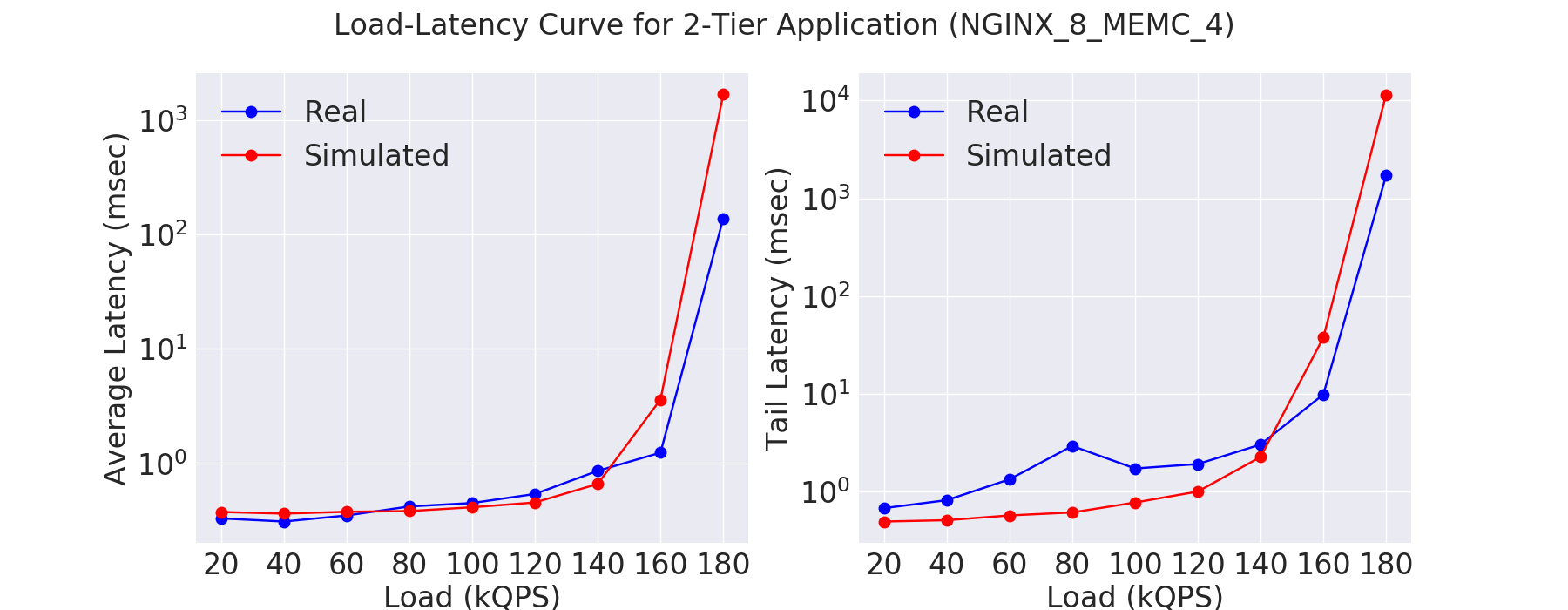} & 
	  \includegraphics[width=0.5\linewidth]{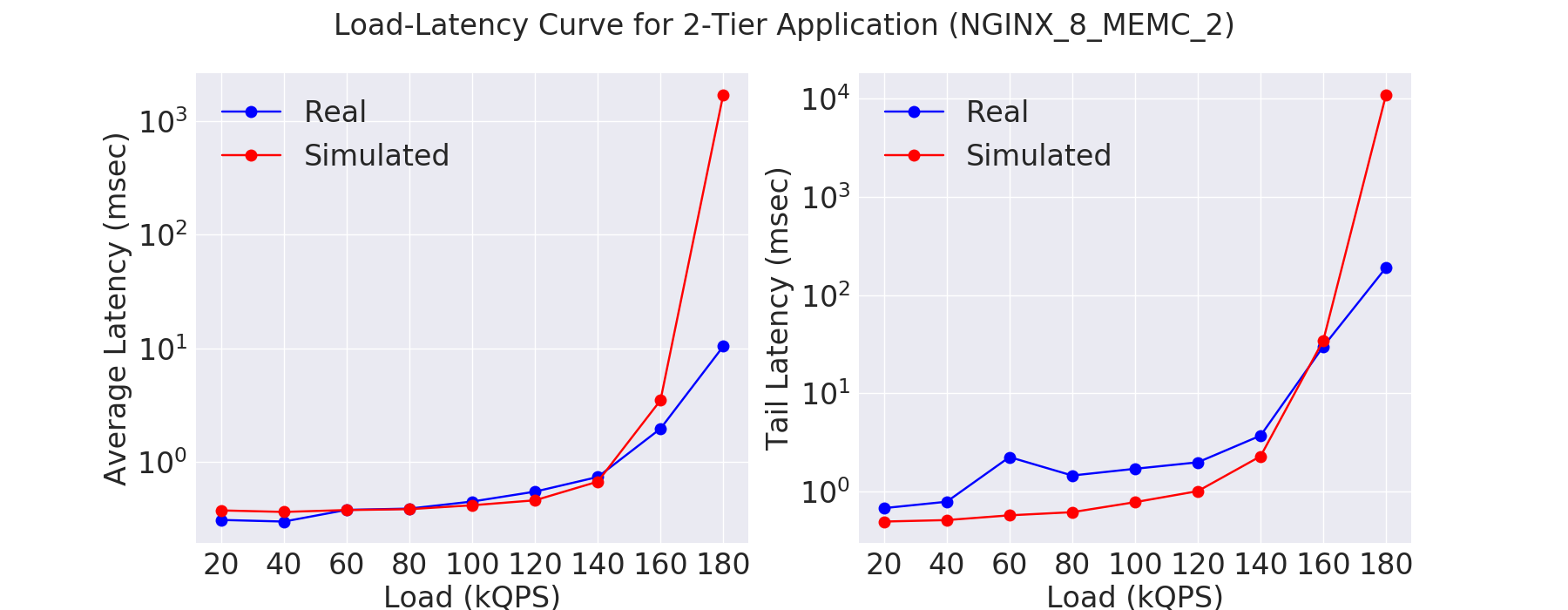} \\ 
	  \includegraphics[width=0.5\linewidth]{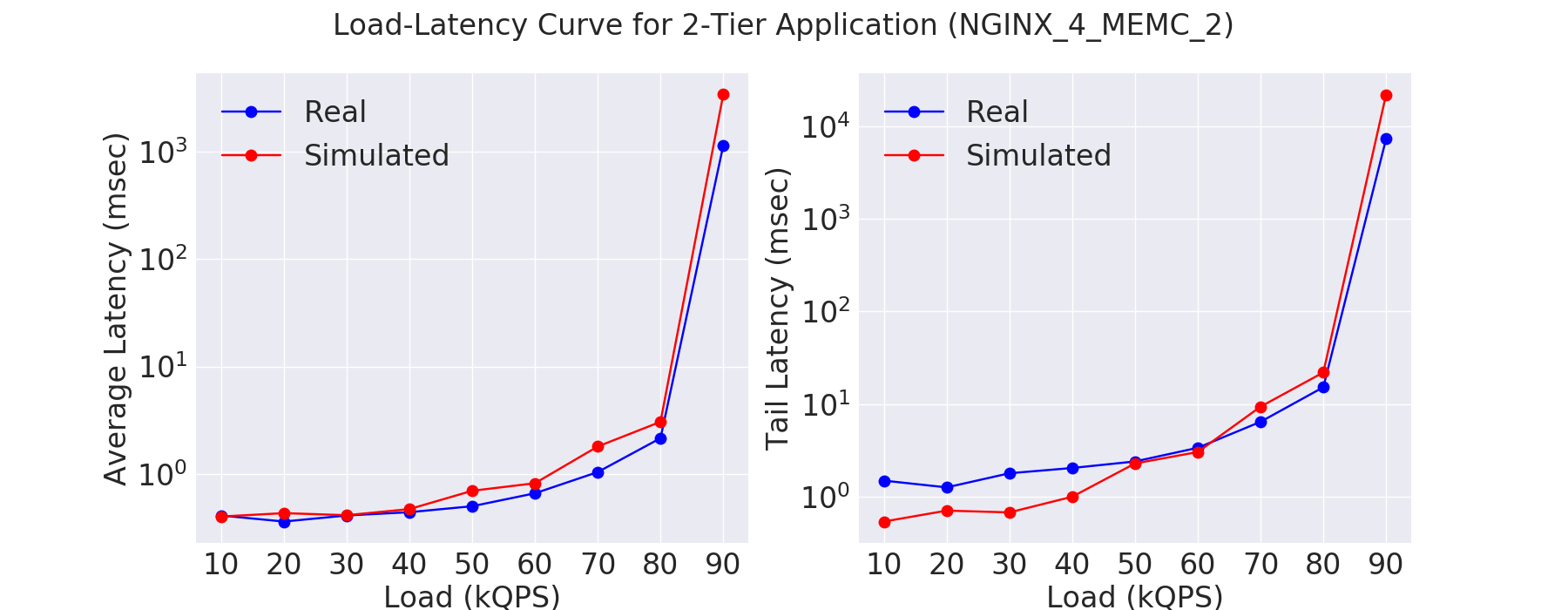} & 
	  \includegraphics[width=0.5\linewidth]{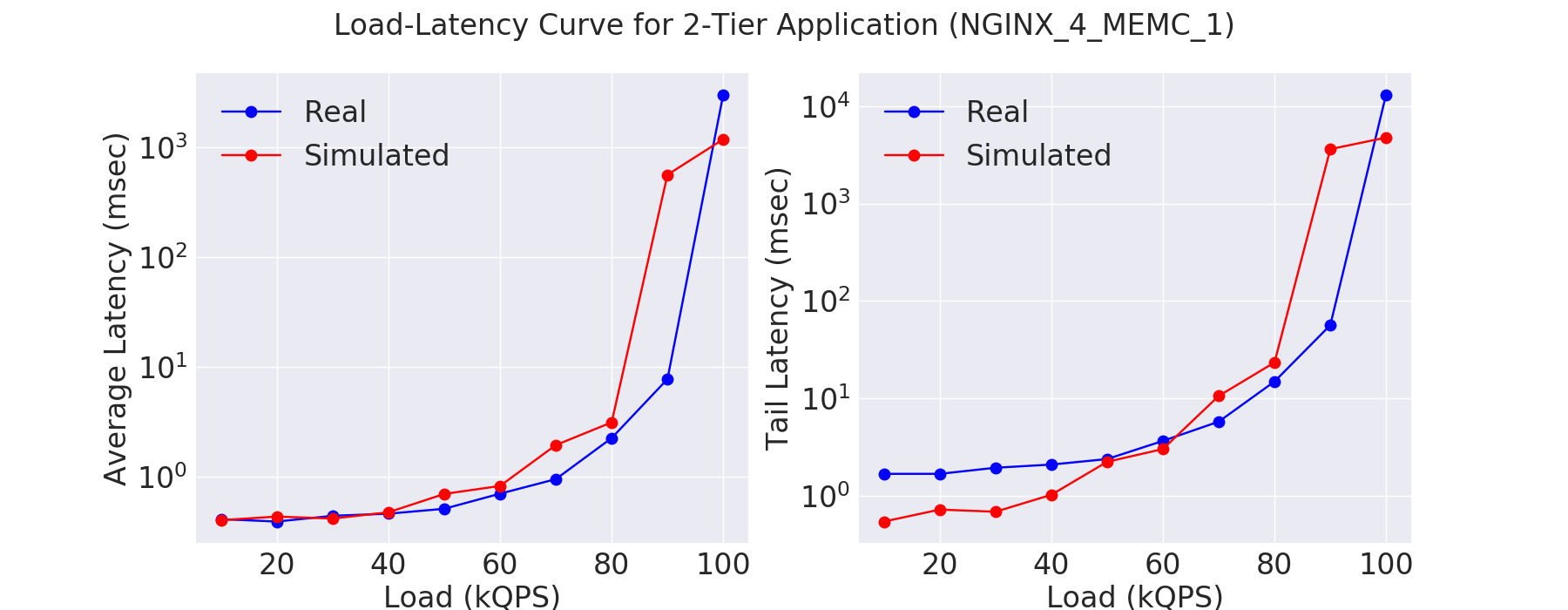}
  \end{tabular}
  \caption{Validation of the two-tier ({\smallcapital NGINX}-memcached) application across different thread configurations for each microservice. }
  \label{fig:2tier_valid}
\end{figure*}

We now validate $\mu$qSim against a set of real microservices running on a server cluster. 
The configuration of each server is detailed in Table~\ref{tab:server_spec}. 
We validate $\mu$qSim with respect to two aspects of application behavior. 
\begin{itemize}
	\item First, we verify that the simulator can reproduce the load-latency curves of real applications, including their saturation point. Given that latency increases exponentially beyond 
		saturation, ensuring that the simulator captures the bottlenecks of the real system is essential in its effectiveness. 
	\item Second, we verify that $\mu$qSim accurately captures the magnitude of the end-to-end average and tail latency of real applications. 
\end{itemize}
		

\subsection{Simple Multi-tier Microservices}

We first validate $\mu$qSim against a simple 2- and 3-tier application, comprised of popular microservices deployed in many production systems. 
The 2-tier service consists of a front-end webserver, implemented using {\smallcapital NGINX}, and an in-memory caching key-value store, implemented with memcached. 
The 3-tier application additionally includes a persistent back-end database, implemented using MongoDB. The architectures of the two applications are shown 
in Fig.\ref{fig:2tier_app}(a) and (b) respectively. 
In the 2-tier service, {\smallcapital NGINX} receives the client request over http 1.1, queries memcached for the requested \texttt{key}, and returns the \texttt{<key,value>} pair to the client. 
In the 3-tier service, {\smallcapital NGINX} first queries the cache for the requested \texttt{key} (memcached), 
and if not present, queries the back-end database (MongoDB). Memcached implements a write-allocate policy; 
on a (mem)cache miss, the \texttt{<key,value>} is also written to memcached to speed up subsequent accesses. 

For all experiments, we use an open-loop workload generator, implemented by modifying the {\texttt{wrk2}} client~\cite{chen2017workload}. The client runs 
on a dedicated server, and uses 16 threads and 320 connections to ensure no client-side saturation. 
For this experiment both job inter-arrival times and request value sizes are exponentially distributed. 
Finally, for both the 2- and 3-tier services, memcached is allocated 1GB memory, and MongoDB has unlimited disk capacity. 


\begin{figure}
  \includegraphics[width=\linewidth]{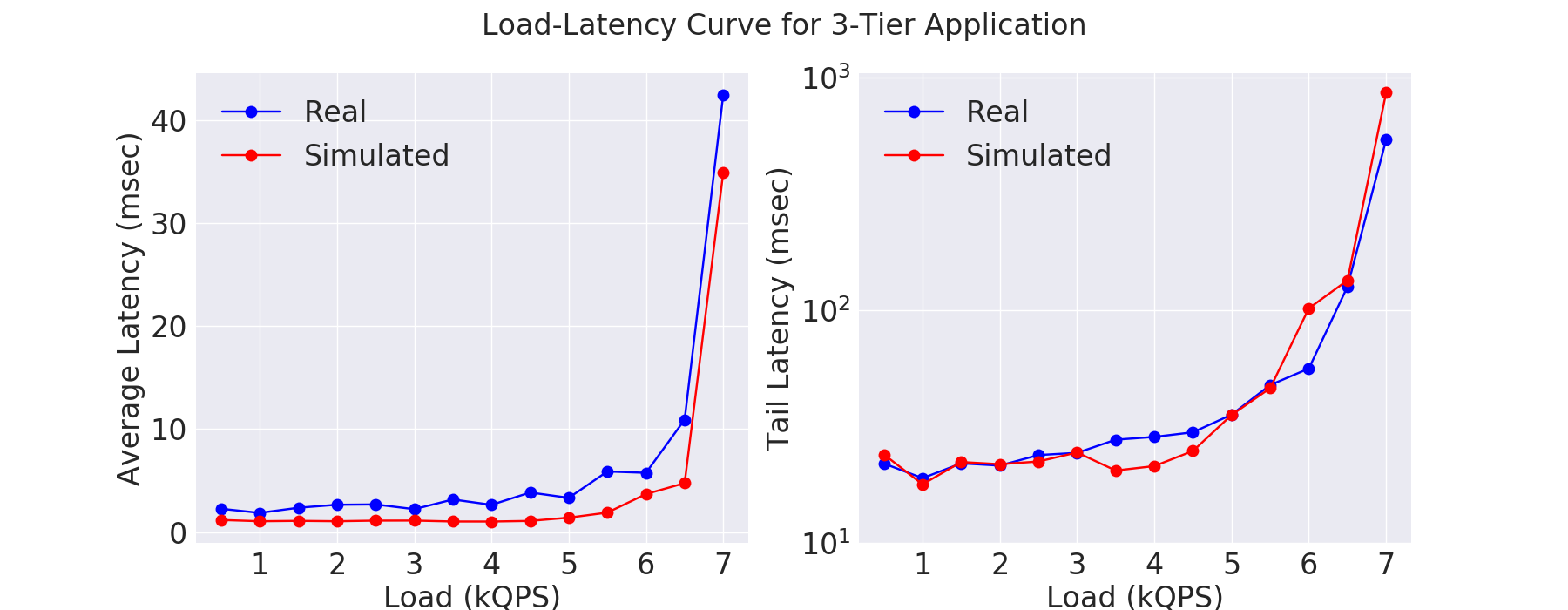}
  \caption{Validation of the three-tier ({\smallcapital NGINX}-memcached-MongoDB) application. }
  \label{fig:3tier_valid}
\end{figure}

For the 2-tier application, we varied the number of threads and processes for {\smallcapital NGINX} and memcached to observe their respective scalability. 
We specifically evaluate configurations with 8 processes for {\smallcapital NGINX} and \{4,2\} threads for memcached, 
as well as a 4-process configuration for {\smallcapital NGINX} and \{2,1\}-thread configurations for memcached. 
The 3-tier application is primarily bottlenecked by the disk I/O bandwidth of MongoDB, so scaling the number 
of downstream microservices does not have a significant impact on performance. We evaluate an 8-process configuration for {\smallcapital NGINX} and 
and a 2-thread configuration for memcached. 
Each thread (or process) of every microservice is pinned to a dedicated physical core to avoid interference from the OS scheduler's decisions. Results reported for the 
real experiments are averaged across 3 runs. 

Fig.~\ref{fig:2tier_valid} shows the comparison between the real and simulated two-tier application across thread/process configurations. 
Across all concurrency settings, $\mu$qSim faithfully reproduces the load-latency curve of the real system, 
including its saturation point. It also accurately captures that giving more resources to memcached does not further improve 
throughput before saturation, since the limiting factor is {\smallcapital NGINX}. 
Additionally, before the 2-tier application reaches saturation, 
the simulated mean latencies are on average 0.17ms away from the real experiments, 
and the simulated tail latencies are on average 0.83ms away from the real ones. 

Fig.~\ref{fig:3tier_valid} shows the same experiment for the 3-tier service. 
The results are again consistent between real and simulated performance, with the 
simulated mean latencies being on average 1.55ms away from real measurements, 
and the simulated tail latencies deviating by 2.32ms on average. 

\subsection{Capturing Load Balancing \& Fanout Effects} 

\noindent{\bf{Load balancing: }}Load balancers are used in most large-scale cloud environments to fairly divide the load across instances of a scale-out application. 
We now examine how accurate $\mu$qSim is in capturing the performance impact of load balancing. We construct 
a load balancing scenario using an instance of {\smallcapital NGINX} as a load-balancing proxy, 
and several instances of {\smallcapital NGINX} of the same setup as the scaled-out webservers behind the proxy. 
For each client request, the proxy chooses one webserver to forward the request to, in a round-robin fashion. 
Fig.~\ref{fig:ngx_loadbalancing_diagram} shows the setup for load balancing, and Fig.~\ref{fig:ngx_lb} shows 
the load-latency (99$^{th}$ percentile) curves for the real and simulated system. The saturation load scales linearly 
for a scale out factor of 4 and 8 from 35kQPS to 70kQPS, and sub-linearly beyond that, e.g., for scale-out of 16, 
saturation happens at 120kQPS as the cores handling the interrupts (\texttt{soft\_irq} processes) saturate before the {\smallcapital NGINX} instances. 
In all cases, the simulator accurately captures the saturation pattern of the real load balancing scenario. 

\noindent{\bf{Request fanout: }}We also experiment with request fanout, which is common in applications with distributed state. 
In this case, a request only completes when responses from all fanout services have been received~\cite{tailatscale}. Request fanout 
is a well-documented source of unpredictable performance in cloud infrastructures, as a single slow leaf node can degrade 
the performance of the majority of user requests~\cite{tailatscale,Delimitrou15,Lo14,Adrenaline}. As with load-balancing, the fanout experiment 
uses an {\smallcapital NGINX} instance as a proxy, which - unlike load balancing - now forwards each request 
to all {\smallcapital NGINX} instances of the next tier. We scale the fanout factor from 4 to 16 servers, and assign 1 core and 1 thread to 
each fanout service. We also dedicate 4 cores to network interrupts. Each requested webpage is 612 bytes in size, 
and the workload generator is set up in a similar way to the 2- and 3-tier experiments above. 
The system configuration is shown in Fig.~\ref{fig:ngx_fanout_diagram} and the load-latency (99$^{th}$ percentile) curve 
for the real and simulated system is shown in Fig.~\ref{fig:ngx_lb}. 
For all fanout configurations, $\mu$qSim accurately reproduces the tail latency and saturation point of the real system, including the fact that as fanout increases, 
there is a small decrease in saturation load, as the probability that a single slow server will degrade the end-to-end tail latency increases.


\subsection{Simulating RPC Requests}

Remote Procedure Calls (RPC) are widely deployed in microservices as a cross-microservice RESTful API. In this section we demonstrate that $\mu$qSim can accurately 
capture the performance of a popular {\smallcapital RPC} framework, Apache Thrift~\cite{thrift}, 
and in the next section we show that it can faithfully reproduce the behavior of complex microservices using Thrift as their APIs. 
We set up a simple Thrift client and server; the server responds with a ``Hello World'' message to each request. 
Given the lack of application logic in this case, all time goes towards processing the {\smallcapital RPC} request. 
The real and simulated results are shown in Fig.\ref{fig:thrift_valid}. 
In both cases the Thrift server saturates beyond 50kQPS, while the low-load latency does not exceed 100us. 
Beyond the saturation point the simulator expects latency that increases more gradually with load compared to the real system. 
The reason for this is that the simulator does not capture timeouts and the associated overhead of reconnections, which can cause the real system's latency 
to increase rapidly. 

\begin{figure}
	\centering
  \includegraphics[width=0.56\linewidth,bb=80 0 380 380]{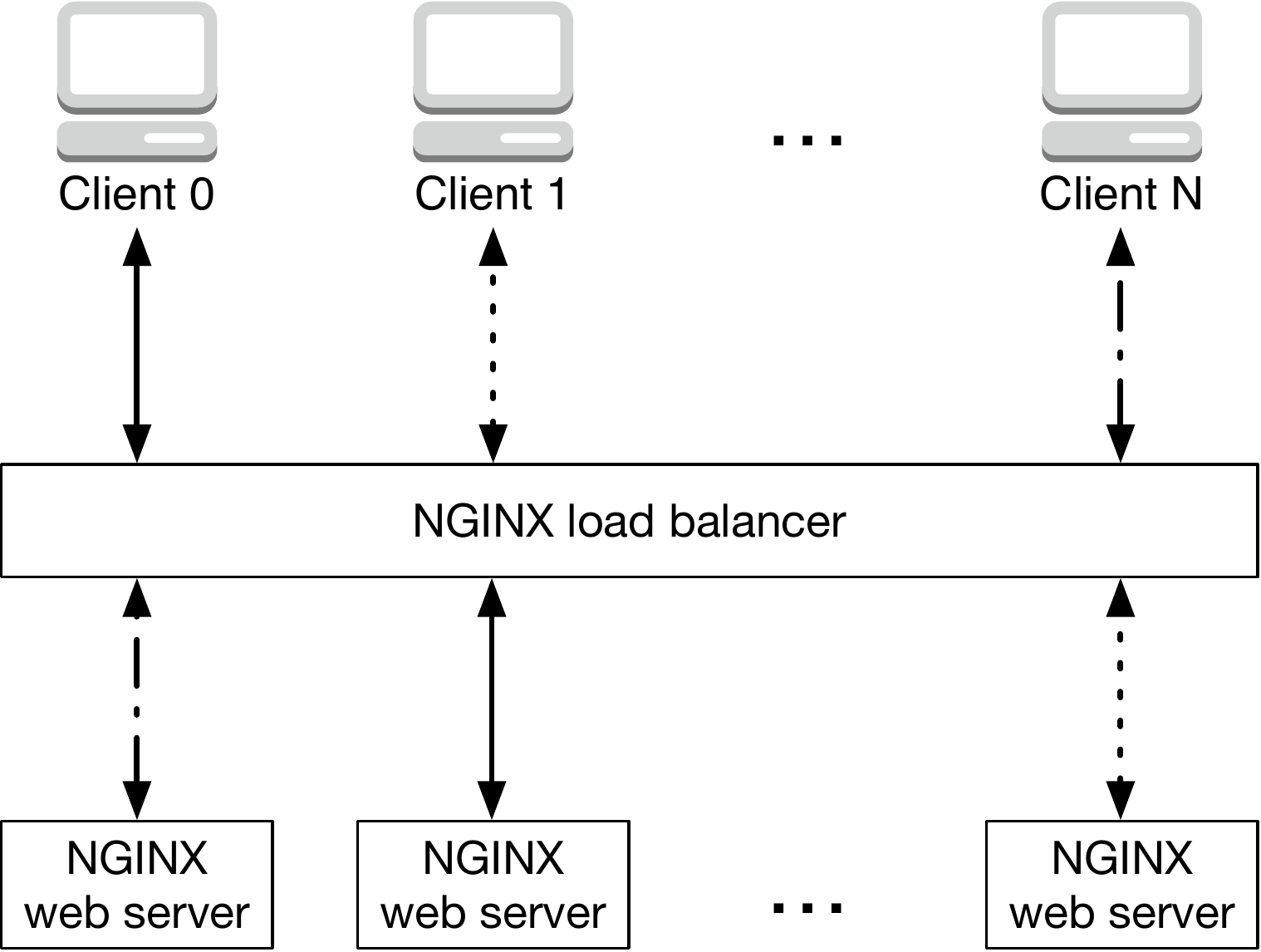}
  \caption{Load balancing in {\smallcapital NGINX}. }
  \label{fig:ngx_loadbalancing_diagram}
\end{figure}

\begin{figure}
  \includegraphics[scale=0.2]{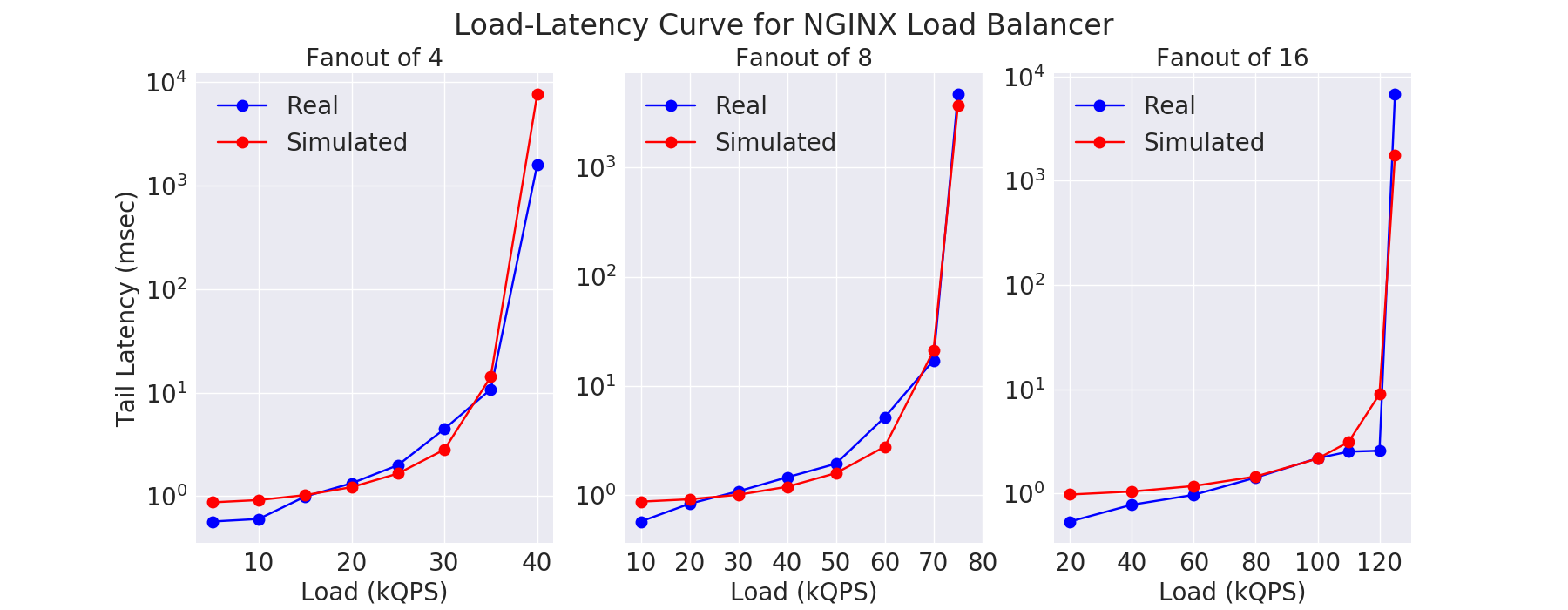}
  \caption{Validation of load balancing results. }
  \label{fig:ngx_lb}
\end{figure}

\subsection{Simulating Complex Microservices}

We have also built a simplified end-to-end application implementing a social network using microservices, 
illustrated in Fig.\ref{fig:thrift_uservice_diagram}. The service implements a unidirectional, broadcast-style social network, where users can follow each other, 
post messages, reply publicly or privately to another user, and browse information about a given user. We focus on the later function in this scenario for simplicity. 
Specifically, the client wants to retrieve a post from a certain user, 
via the \textit{Thrift Frontend} by specifying a given \texttt{userId} and \texttt{postId}. Upon receiving this request from the client, the \textit{Thrift Frontend} 
sends requests to \textit{User Service} and \textit{Post Service}, which search for the user profile and corresponding post respectively. 
Once the user and post information are received, \textit{Thrift Frontend} extracts any media embedded to the user's post via \textit{Media Service}, composes a response 
with the user's metadata, post content, and media (if applicable), and returns the response to the client. The user, post, and media objects are stored in the corresponding 
MongoDB instances, and cached in memcached to lower request latency. All cross-microservice communication in this application happens using Apache Thrift. 
The comparison between the real and simulated system is shown in Fig.~\ref{fig:thrift_valid}b. At low load, $\mu$qSim closely matches the latency of the real application, while at high load 
it saturates at a similar throughput as the real social network service. 
This application contains a large number of queues and dependent microservices, including applications with fanout, synchronization, and blocking characteristics, showing that $\mu$qSim can capture 
the behavior of complex microservices accurately. 



\begin{figure}
  \centering
  \includegraphics[width=0.6\linewidth,bb=60 0 360 300]{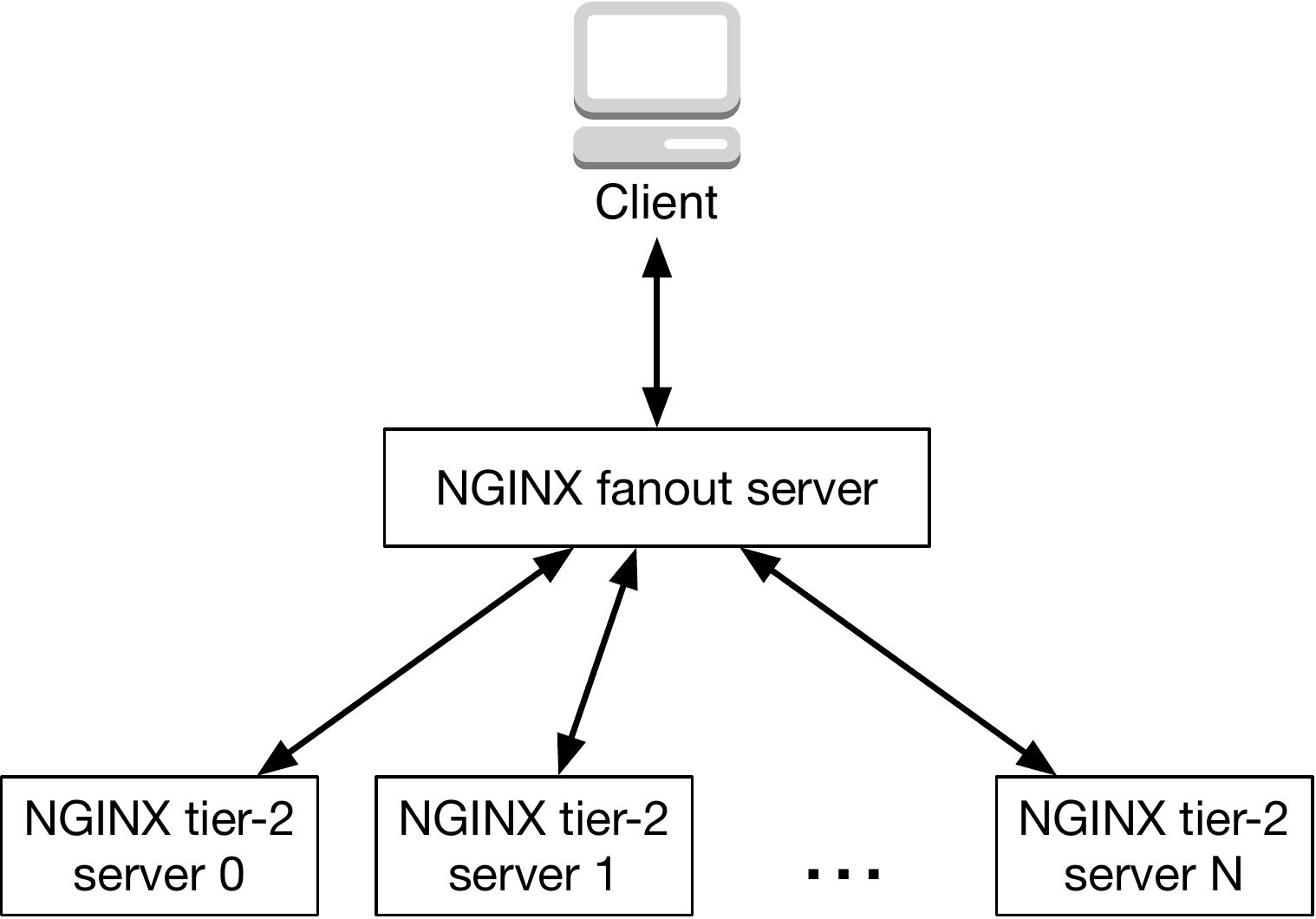}
  \caption{Request fanout in {\smallcapital NGINX}. }
  \label{fig:ngx_fanout_diagram}
\end{figure}

\begin{figure}
  \includegraphics[scale=0.225]{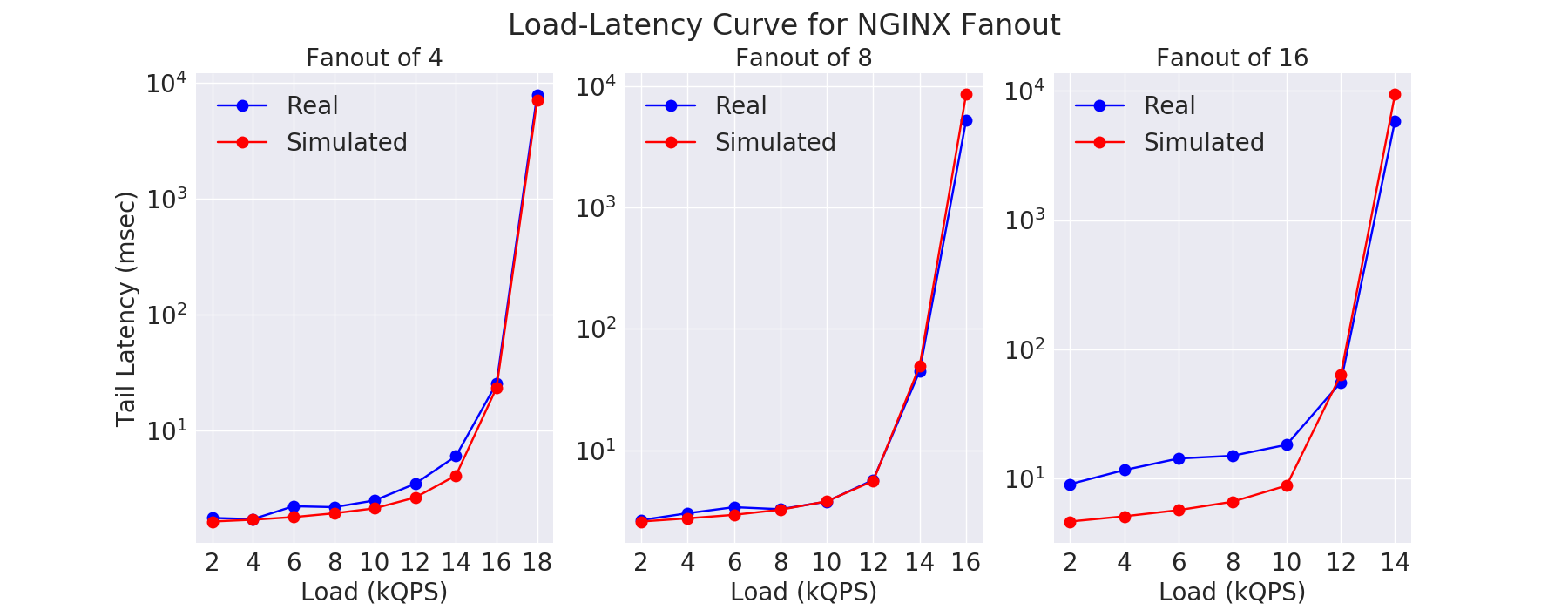}
  \caption{Validation of request fanout impact in {\smallcapital NGINX}. }
  \label{fig:ngx_fanout}
\end{figure}

\subsection{Comparison with BigHouse}

Fig.~\ref{fig:bighouse_compare} shows the comparison of $\mu$qSim and BigHouse simulating a single-process {\smallcapital NGINX} webserver and a 4-thread memcached. 
In BigHouse both {\smallcapital NGINX} and memcached are modeled as a single stage server. 
In $\mu$qSim we use the model shown in Fig.~\ref{lst:memc_json} for memcached, and we adopt a similar model for {\smallcapital NGINX}, 
consisting of two stages: \texttt{epoll} and \texttt{handler\_processing}.
For both applications, $\mu$qSim captures the real saturation point closely, while BigHouse saturates at much lower load than the real experiments. 
The reason is that in $\mu$qSim the processing time of batching stage \texttt{epoll} is amortized across all batched requests, as in the real system. 
In BigHouse, however, each application is modeled as a single stage so the entire processing time of \texttt{epoll} is accounted for in every request, 
leading to overestimation of the accumulated tail latency.

\begin{figure}
  \centering
  \includegraphics[width=0.6\linewidth,bb=80 0 380 380]{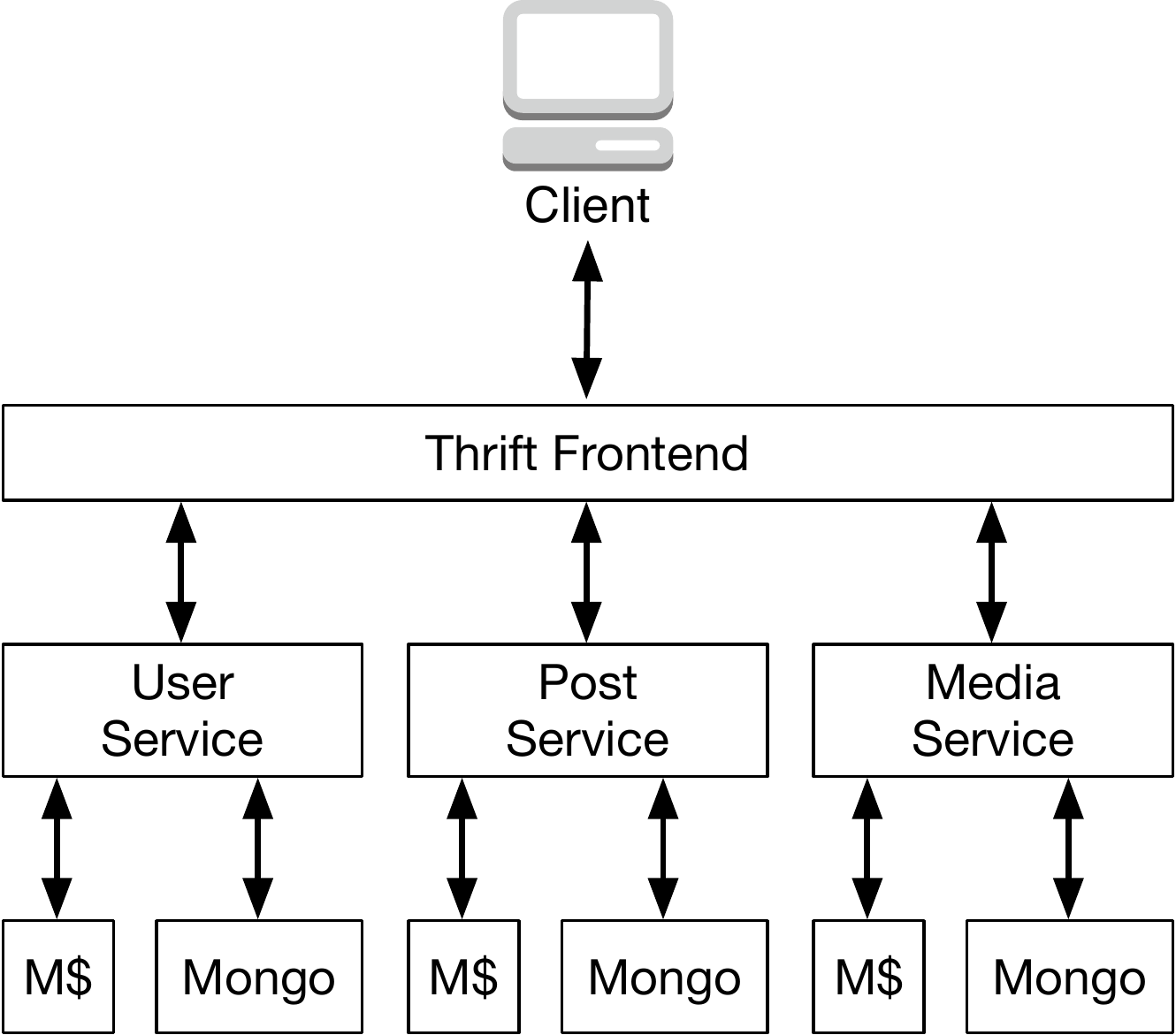}
  \caption{Architecture of the social network microservices application. }
  \label{fig:thrift_uservice_diagram}
\end{figure}

\begin{figure}
  \centering
  \begin{subfigure}[b]{0.47\linewidth}
    \includegraphics[width=\linewidth]{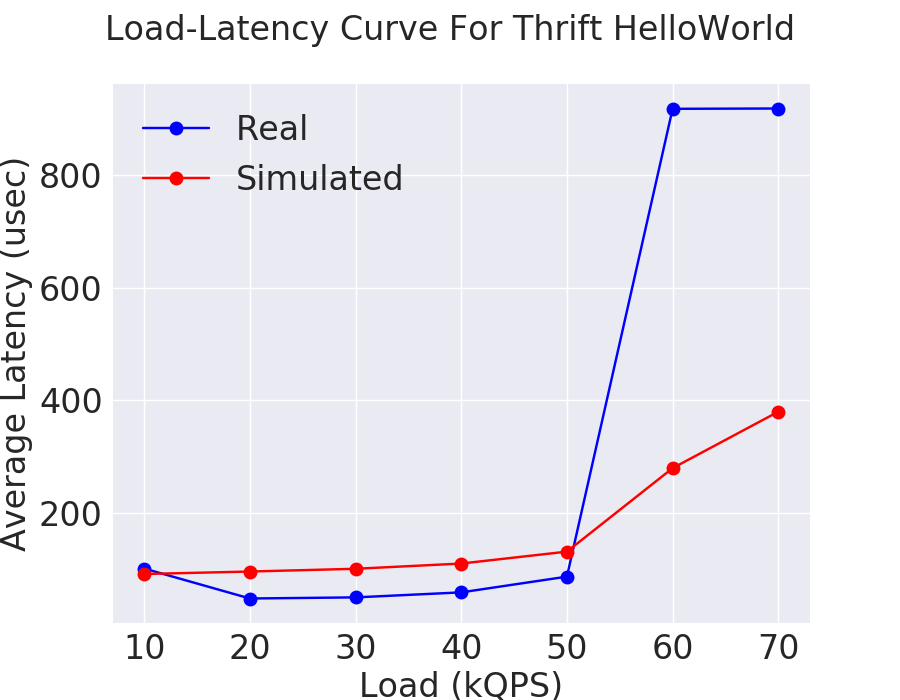}
  \end{subfigure}
  \begin{subfigure}[b]{0.47\linewidth}
    \includegraphics[width=\linewidth]{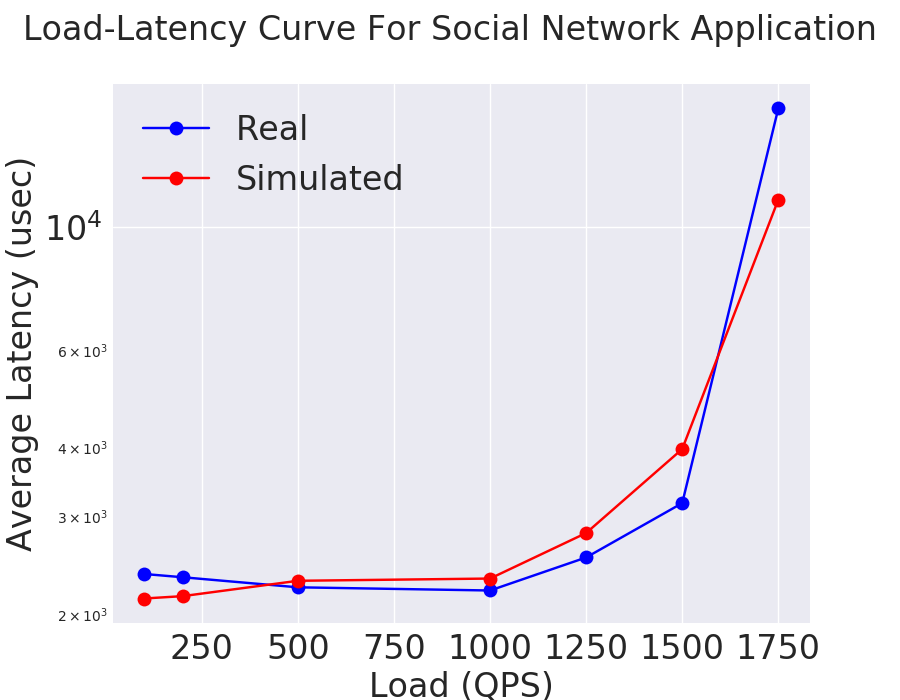}
  \end{subfigure}
  \caption{Validation of RPC request simulation using Apache Thrift~\cite{thrift}. }
  \label{fig:thrift_valid}
\end{figure}

\begin{figure}
  \centering
  \begin{subfigure}[b]{0.85\linewidth}
    \includegraphics[width=\linewidth]{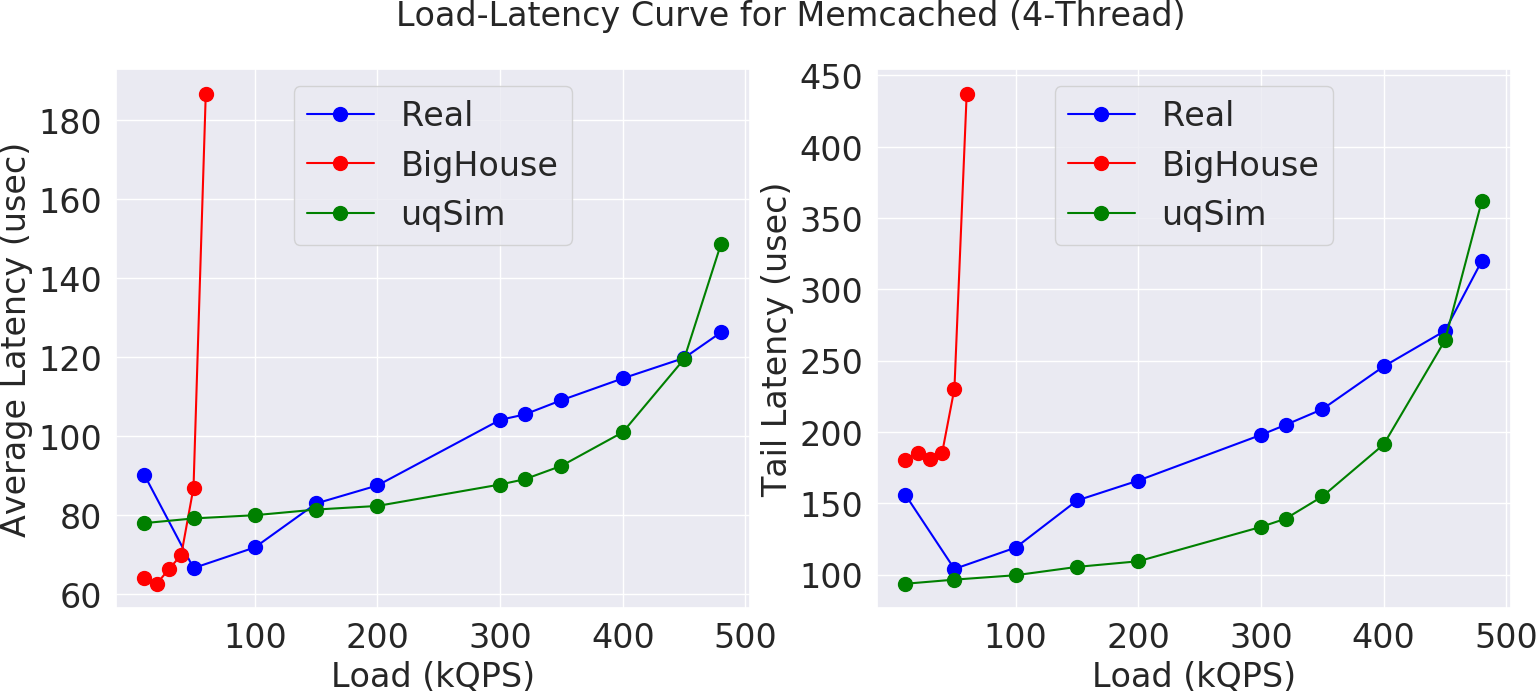}
  \end{subfigure}
  \begin{subfigure}[b]{0.85\linewidth}
    \includegraphics[width=\linewidth]{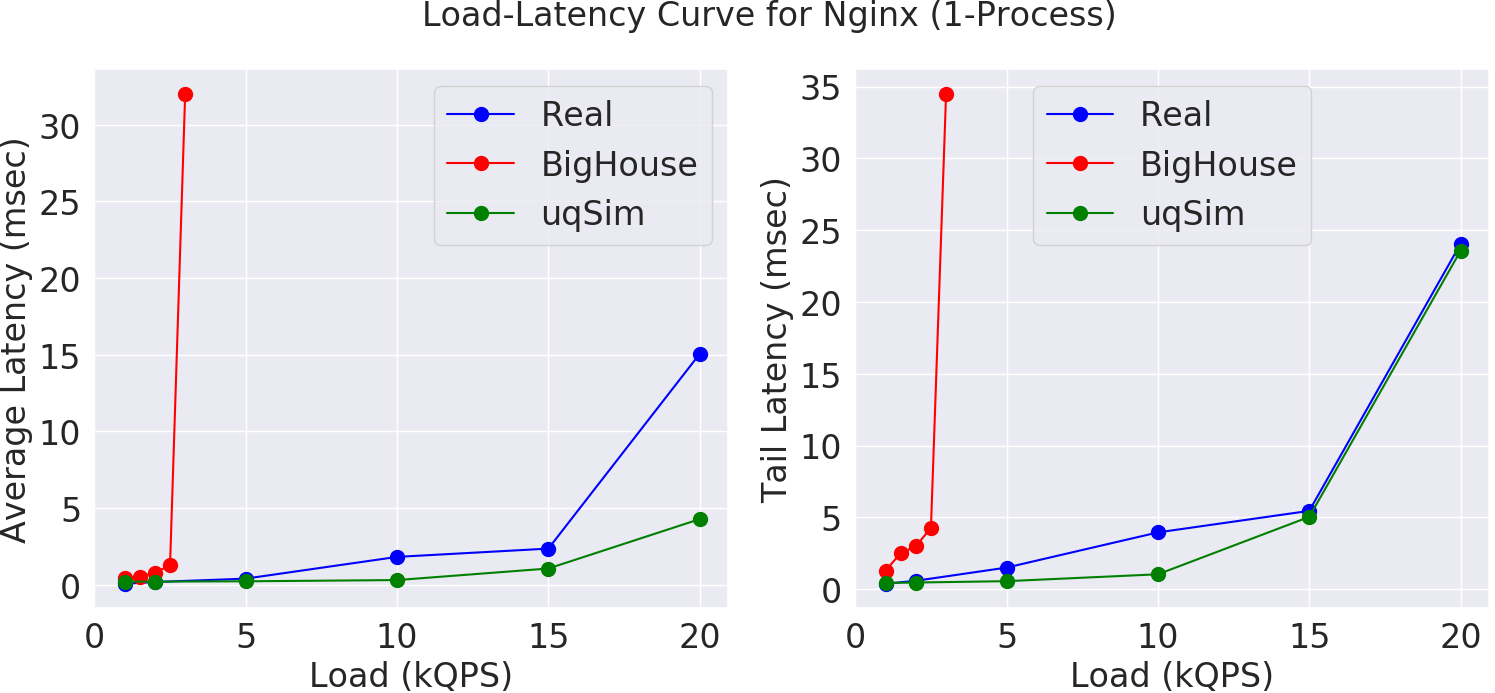}
  \end{subfigure}
  \caption{Comparison of $\mu$qSim and BigHouse}
  \label{fig:bighouse_compare}
\end{figure}

\section{Use Cases}

In this section we discuss two case studies that leverage $\mu$qSim to obtain performance and efficiency benefits. 
In the first case study, we experiment with the effect of slow servers in clusters of different sizes 
in order to reproduce the tail-at-scale effects documented in~\cite{tailatscale}.
In the second study, we design a power management algorithm for interactive microservices, 
and show its behavior on real and simulated servers. The platforms we use for real experiments are the same as before (Table~\ref{tab:server_spec}). 

\subsection{Tail@Scale Effect}

As cluster sizes and request fanouts grow, the impact of a small fraction of slow machines are amplified, since a few stragglers dominate tail latency. 
In this scenario we simulate clusters of different sizes, ranging from 5 servers to 1000 servers, 
in a similar fanout configuration as the one discussed in~\cite{tailatscale}. Under this setup 
a user request fans out to all servers in the cluster, and only 
returns to the user after the last server responds. To capture similar effects as in~\cite{tailatscale} 
the application is a simple one-stage queueing system with exponentially distributed processing time, around a 1ms mean. 
To emulate slow servers, we increase the average processing time of a configurable fraction of randomly-selected servers by $10\times$. 
Fig.~\ref{fig:tail_at_scale} shows the impact of slow servers on tail latency as fanout (cluster size) increases. 
For the same percentage of slow servers, e.g., 1\%, the larger the size of the cluster, 
the more likely it is that tail latency is defined by the slow machines. 
Similarly, as the fraction of slow servers increases, so does the probability for high 
tail latency, complying to the probabilistic expression discussed in~\cite{tailatscale}. 
Note that for cluster sizes greater than 100 servers, 1\% of slow servers is sufficient 
to drive tail latency high, consistent with the results in~\cite{tailatscale}. 


\begin{figure}
	\centering
  \includegraphics[width=0.8\linewidth]{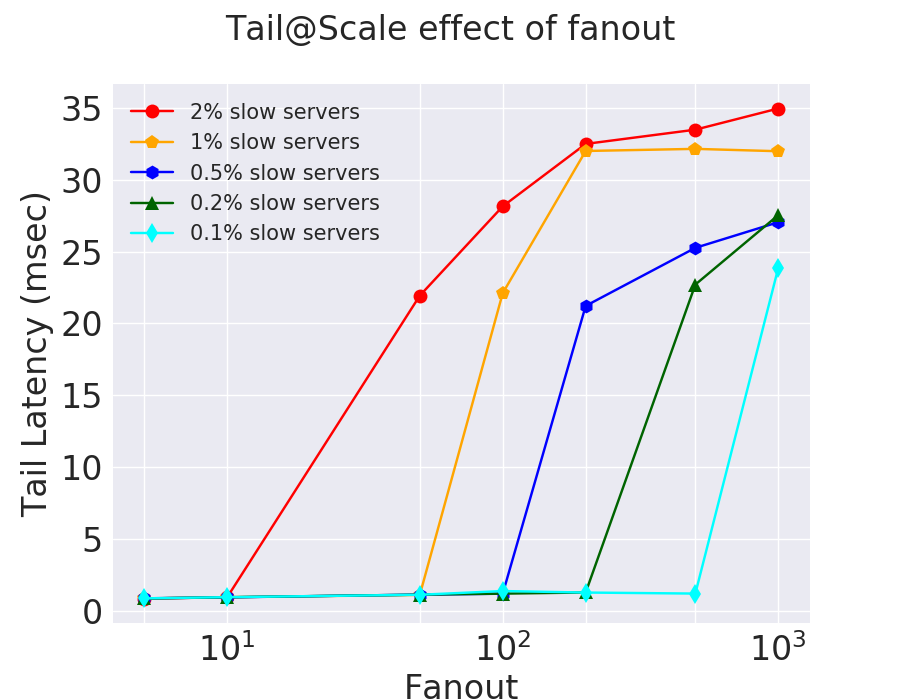}
  \caption{The tail at scale effects of request fanout. }
  \label{fig:tail_at_scale}
\end{figure}


\begin{algorithm}
\begin{algorithmic}[1]
\caption{Power Management Algorithm}
\label{alg:pm_algo}
\While{$True$}
    \If{$time_{now} - time_{prev\_cycle} \textless Interval$}
        \State sleep(Interval)
    \EndIf
    \If{$stats[end2end]$ $<$ $Target$}
        \If{$stats\ no\_relaxed\_than\ fail\_tuples$ }
          \State bucket.insert(stats)
        \EndIf
        \State increase bucket.preference 
  \If{$Cycle Count$ $>$ $Interval$ }
            \State Choose new target bucket
            \State Choose per-tier QoS
        \EndIf
        \State Slow down at most 1 tier based on slack
    \Else
        \State decrease bucket.preference
        \State bucket.failing\_list.insert(current\ target)
        \State Choose new target bucket 
        \State Choose per-tier QoS
        \State Speed up all tiers with higher latency than target
    \EndIf
\EndWhile
\end{algorithmic}
\end{algorithm}

\subsection{Power Management}

The lack of energy proportionality in datacenters is a well-documented problem~\cite{meisner09,Lo14,Lo15}. 
Quality-of-service-aware power management is challenging for cloud applications, 
as resource utilization does not always correlate closely with QoS, resulting in significant increases in tail latency, 
even when the server is not saturated~\cite{Lo14}. Power management is even more challenging 
in multi-tier applications and microservices, since dependencies between neighboring microservices 
introduce backpressure effects, creating cascading hotspots and QoS violations through the system. This makes it hard 
to determine the appropriate frequency setting for each microservice, and to identify which microservice is the culprit of a QoS violation. 

Our proposed power management algorithm is based on the intuition that reasoning about QoS guarantees for the end-to-end application 
requires understanding the interactions between dependent microservices, and how changing the performance requirements of one tier 
affects the rest of the application. We adopt a divide-and-conquer approach by dividing the end-to-end QoS requirement to per-tier QoS requirements,
because, as queueing theory suggests, the combination of per-tier state should be recurrent and reproducible, 
which indicates that as long as the per-tier latencies achieve values that have allowed the end-to-end QoS to be met in the past, 
the system should be able to recover from a QoS violation. 


Based on this intuition, our algorithm divides the tail latency space into a number of buckets, with each bucket corresponding to a given end-to-end QoS range, and classifies 
the observed per-tier latencies into the corresponding buckets. At runtime, the scheduler picks one per-tier latency tuple from a certain bucket, and uses it as the per-tier QoS target.
Different buckets are equally likely to be visited initially, and as the application execution progresses, the scheduler learns which buckets are more likely to meet the end-to-end tail latency 
requirement, and adjusts the weights accordingly. To refine the recorded per-tier latencies, every bucket also keeps a list of previous per-tier tuples 
that fail to meet QoS when used as the latency target, and a new per-tier tuple is only inserted if it is no more relaxed than any of the failing tuples of the corresponding bucket.
This way the scheduler eventually converges to a set of per-tier QoS requirements that have a high probability to meet the required end-to-end performance target. 
In order to test whether more aggressive power management settings are acceptable, 
the scheduler periodically selects a tier with high latency slack to slow down, and observes the change in end-to-end performance. 
The scheduler only slows down 1 tier at a time, to prevent cascading violations caused by interactions between tiers (like connection pool exhaustion and blocking).
The pseudo-code for the power management algorithm is shown in Algo.~\ref{alg:pm_algo}. 

\begin{figure}
  \includegraphics[width=\linewidth]{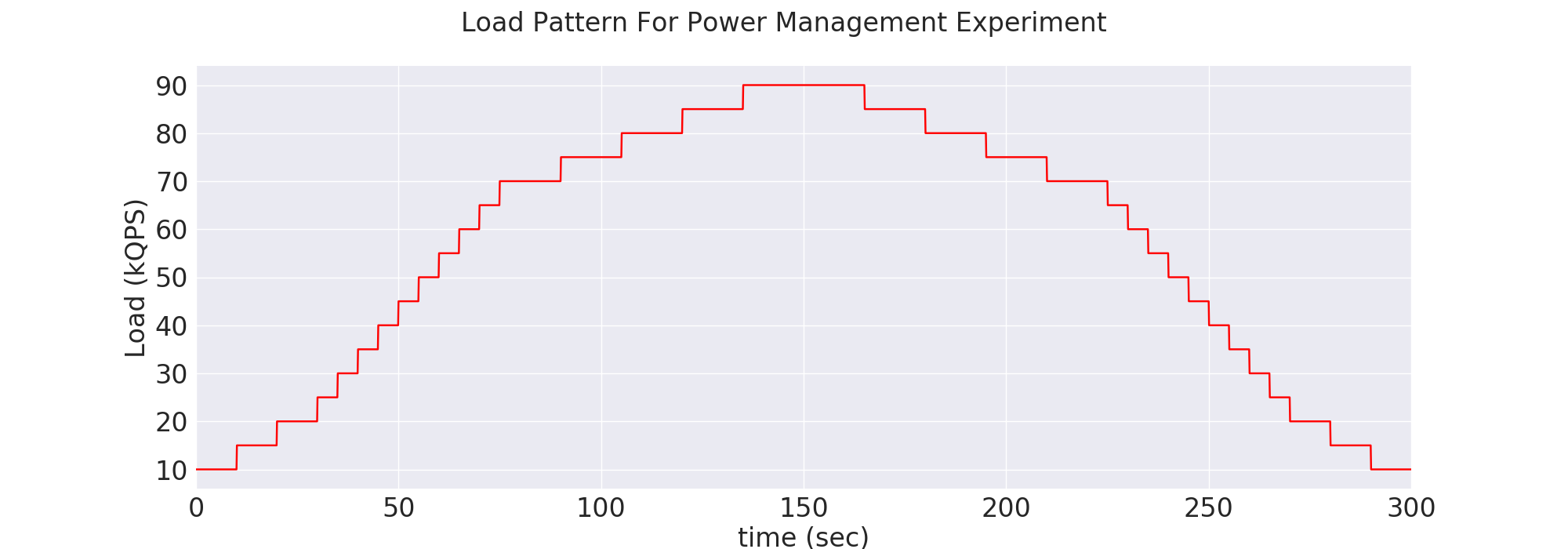}
  \caption{Load fluctuation under the examined diurnal load. }
  \label{fig:diurnal_pattern}
\end{figure}

\begin{table}
	\vspace{0.23in}
\centering
\caption{Power management QoS violation rates. } 
\label{tab:violation_rate}
	\begin{tabular}{lllllll}
		\cline{1-4}
		\cline{1-4}
		\hline
		\multicolumn{1}{|l|}{Decision Intervals} & \multicolumn{1}{l|}{0.1s}  & \multicolumn{1}{l|}{0.5s}  & \multicolumn{1}{l|}{1s}          \\ \hline
		\hline
		\multicolumn{1}{|l|}{Simulated System} & \multicolumn{1}{l|}{0.6\%} & \multicolumn{1}{l|}{2.2\%} & \multicolumn{1}{l|}{5.0\%}    \\ 
		\multicolumn{1}{|l|}{Real System}      & \multicolumn{1}{l|}{1.5\%} & \multicolumn{1}{l|}{2.7\%} & \multicolumn{1}{l|}{6.0\%}   \\ \hline
		\hline
	\end{tabular}
\end{table}

\begin{figure*}
  \centering
  \begin{subfigure}[b]{\linewidth}
    \includegraphics[scale=0.29, viewport=170 0 0 288]{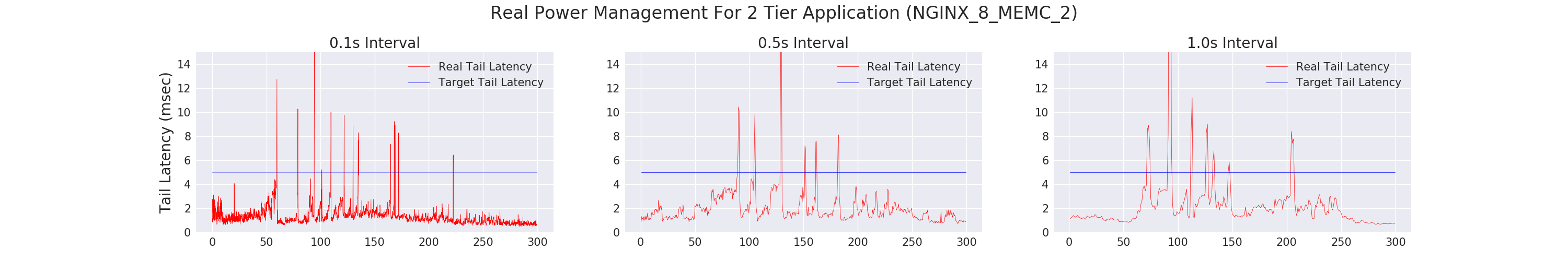}
  \end{subfigure}
  \begin{subfigure}[b]{\linewidth}
    \includegraphics[scale=0.29, viewport=170 0 0 720]{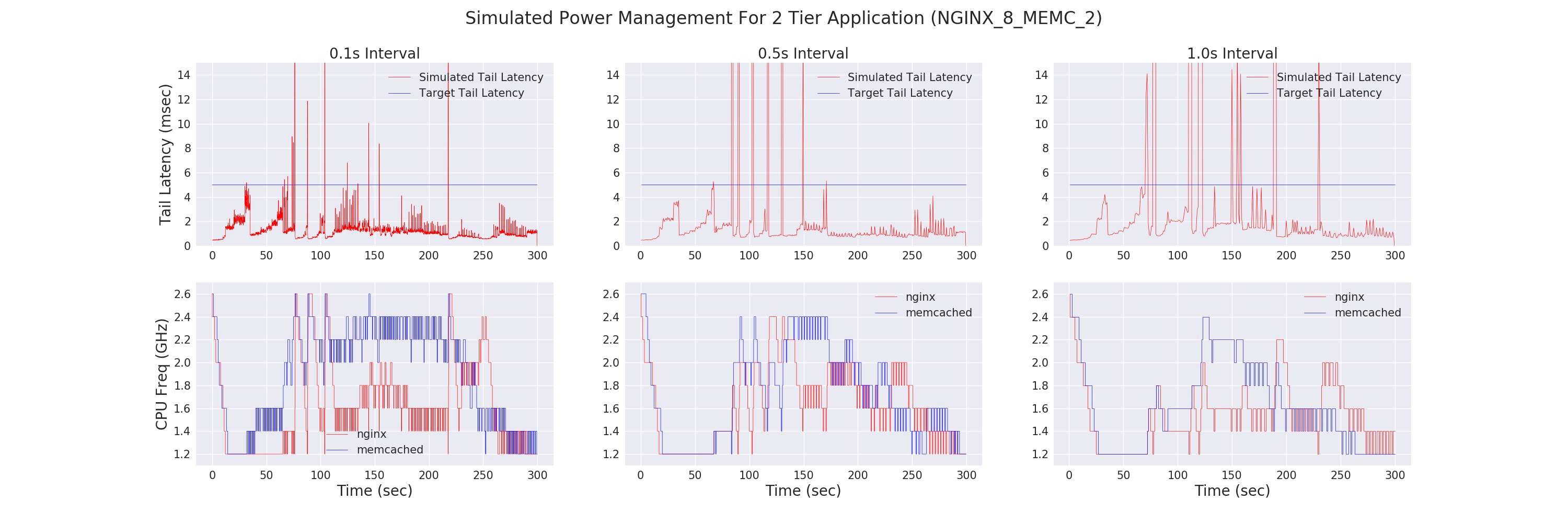}
  \end{subfigure}
  \caption{Tail latency and frequency settings when employing the power management mechanism of Algo.~\ref{alg:pm_algo}. We show tail latencies for both the real and simulated systems. }
  \label{fig:power_management}
\end{figure*}

We evaluate the power management algorithm above with the 2-tier application both using 
$\mu$qSim and real servers. To highlight the potential of power management, we drive the application with a diurnal input load, shown in Fig.~\ref{fig:diurnal_pattern}. 
To simulate the impact of power management in $\mu$qSim, we adjust the processing time of each execution stage as frequency changes 
by providing histograms corresponding to different frequencies. 
We also vary the decision interval of the scheduler, from $0.1s$ to $1s$. Fig.\ref{fig:power_management} shows the tail latency 
and per-tier frequency over time under different decision intervals, and Table~\ref{tab:violation_rate} shows the fraction of time for 
which QoS is violated. 

Unsurprisingly, the real system is slightly more noisy compared to $\mu$qSim, due 
to effects not modeled in the simulator, such as request timeouts, {\smallcapital TCP/IP} contention, 
and operating system interference from scheduling and context switching. 
The lower latency jitter in the simulator also results in less frequent changes in power management decisions. 
Nonetheless, both systems follow similar patterns in their decision process, 
and converge to very similar tail latencies. The reason why tail latency in both cases 
converges to around 2ms despite a 5ms QoS target, is the coarse frequency-voltage granularity 
of the power management technique we use -- {\smallcapital DVFS}. 
The discrete frequency settings enabled by {\smallcapital DVFS} can only lead to discrete processing speeds, and therefore discrete latency ranges.
As a result, further lowering frequency to reduce latency slack would result in QoS violations. 
More fine-grained power management techniques, such as {\smallcapital RAPL}~\cite{Lo14}, would help bring the instantaneous tail latency closer to the QoS. 


\section{Conclusions}
\label{sec:conclusions}

We presented $\mu$qSim, a scalable and validated queueing network simulator for interactive microservices. 
$\mu$qSim offers detailed models both for execution phases within a single microservice, and across complex dependency graphs 
of microservices. We have validated $\mu$qSim against applications with few up to many tiers, as well as scenarios of load balancing 
and request fanout, and showed minimal differences in throughput and tail latency in all cases. Finally, we showed that $\mu$qSim can be used to 
gain insight into the performance effects that emerge in systems of scale larger than what can be evaluated outside a production cloud environment, 
as well as when using mechanisms, such as power management, that aim to improve the resource efficiency of large-scale datacenters. 
We plan to open-source $\mu$qSim to motivate more work in the field of microservices. 

\begin{center}
{{A\small{CKNOWLEDGEMENTS}}}
\end{center}

We sincerely thank the anonymous reviewers for their feedback 
on earlier versions of this manuscript. This work was supported 
in part by NSF grant CNS-1422088, a Facebook Faculty Research Award, 
a John and Norma Balen Sesquicentennial Faculty Fellowship, and 
generous donations from Google Compute Engine, Windows Azure, and Amazon EC2.

\bibliographystyle{ieeetr}
\balance
\bibliography{references}

\begin{thebibliography}{10}

\bibitem{BarrosoBook}
L.~Barroso and U.~Hoelzle, {\em The Datacenter as a Computer: An Introduction
  to the Design of Warehouse-Scale Machines}.
\newblock MC Publishers, 2009.

\bibitem{Barroso11}
L.~Barroso, ``Warehouse-scale computing: Entering the teenage decade,'' {\em
  ISCA Keynote, SJ, June 2011}.

\bibitem{GoogleTrace}
C.~Reiss, A.~Tumanov, G.~Ganger, R.~Katz, and M.~Kozych, ``Heterogeneity and
  dynamicity of clouds at scale: Google trace analysis,'' in {\em Proceedings
  of SOCC}, 2012.

\bibitem{Lo15}
D.~Lo, L.~Cheng, R.~Govindaraju, P.~Ranganathan, and C.~Kozyrakis, ``Heracles:
  Improving resource efficiency at scale,'' in {\em Proc. of the 42Nd Annual
  International Symposium on Computer Architecture (ISCA)}, Portland, OR, 2015.

\bibitem{Delimitrou13}
C.~Delimitrou and C.~Kozyrakis, ``{Paragon: QoS-Aware Scheduling for
  Heterogeneous Datacenters},'' in {\em Proceedings of the Eighteenth
  International Conference on Architectural Support for Programming Languages
  and Operating Systems (ASPLOS)}, Houston, TX, USA, 2013.

\bibitem{Delimitrou14}
C.~Delimitrou and C.~Kozyrakis, ``{Quasar: Resource-Efficient and QoS-Aware
  Cluster Management},'' in {\em Proc. of ASPLOS}, Salt Lake City, 2014.

\bibitem{Delimitrou14b}
C.~Delimitrou and C.~Kozyrakis, ``{Quality-of-Service-Aware Scheduling in
  Heterogeneous Datacenters with Paragon},'' in {\em IEEE Micro Special Issue
  on Top Picks from the Computer Architecture Conferences}, May/June 2014.

\bibitem{Delimitrou16}
C.~Delimitrou and C.~Kozyrakis, ``{HCloud: Resource-Efficient Provisioning in
  Shared Cloud Systems},'' in {\em Proceedings of the Twenty First
  International Conference on Architectural Support for Programming Languages
  and Operating Systems (ASPLOS)}, April 2016.

\bibitem{Delimitrou17}
C.~Delimitrou and C.~Kozyrakis, ``{Bolt: I Know What You Did Last Summer... In
  The Cloud},'' in {\em Proc. of the Twenty Second International Conference on
  Architectural Support for Programming Languages and Operating Systems
  (ASPLOS)}, 2017.

\bibitem{Delimitrou13e}
C.~Delimitrou, N.~Bambos, and C.~Kozyrakis, ``{QoS-Aware Admission Control in
  Heterogeneous Datacenters},'' in {\em Proceedings of the International
  Conference of Autonomic Computing (ICAC)}, San Jose, CA, USA, 2013.

\bibitem{Mars13b}
J.~Mars and L.~Tang, ``Whare-map: heterogeneity in "homogeneous"
  warehouse-scale computers,'' in {\em Proceedings of ISCA}, Tel-Aviv, Israel,
  2013.

\bibitem{Nathuji10}
R.~Nathuji, A.~Kansal, and A.~Ghaffarkhah, ``Q-clouds: Managing performance
  interference effects for qos-aware clouds,'' in {\em Proceedings of EuroSys},
  Paris,France, 2010.

\bibitem{Nathuji07}
R.~Nathuji, C.~Isci, and E.~Gorbatov, ``Exploiting platform heterogeneity for
  power efficient data centers,'' in {\em Proceedings of ICAC}, Jacksonville,
  FL, 2007.

\bibitem{Mars11a}
J.~Mars, L.~Tang, R.~Hundt, K.~Skadron, and M.~L. Soffa, ``Bubble-up:
  increasing utilization in modern warehouse scale computers via sensible
  co-locations,'' in {\em Proceedings of MICRO}, Porto Alegre, Brazil, 2011.

\bibitem{Mars13a}
H.~Yang, A.~Breslow, J.~Mars, and L.~Tang, ``Bubble-flux: precise online qos
  management for increased utilization in warehouse scale computers,'' in {\em
  Proceedings of ISCA}, 2013.

\bibitem{tailatscale}
J.~Dean and L.~A. Barroso, ``The tail at scale,'' in {\em CACM, Vol. 56 No. 2}.

\bibitem{Delimitrou13d}
C.~Delimitrou and C.~Kozyrakis, ``{QoS-Aware Scheduling in Heterogeneous
  Datacenters with Paragon},'' in {\em ACM Transactions on Computer Systems
  (TOCS), Vol. 31 Issue 4}, December 2013.

\bibitem{Cockroft15}
``Microservices workshop: Why, what, and how to get there.''
  \url{http://www.slideshare.net/adriancockcroft/microservices-workshop-craft-conference}.

\bibitem{Cockroft16}
``The evolution of microservices.''
  \url{https://www.slideshare.net/adriancockcroft/evolution-of-microservices-craft-conference},
  2016.

\bibitem{twitter_decomposing}
``Decomposing twitter: Adventures in service-oriented architecture.''
  {\scriptsize\url{https://www.slideshare.net/InfoQ/decomposing-twitter-adventures-in-serviceoriented-architecture}}.

\bibitem{Ueda16}
T.~Ueda, T.~Nakaike, and M.~Ohara, ``Workload characterization for
  microservices,'' in {\em Proc. of IISWC}, 2016.

\bibitem{Sriraman18}
A.~Sriraman and T.~F. Wenisch, ``usuite: {A} benchmark suite for
  microservices,'' in {\em 2018 {IEEE} International Symposium on Workload
  Characterization, {IISWC} 2018, Raleigh, NC, USA, September 30 - October 2,
  2018}, pp.~1--12, 2018.

\bibitem{Zhou18}
X.~Zhou, X.~Peng, T.~Xie, J.~Sun, C.~Xu, C.~Ji, and W.~Zhao, ``Benchmarking
  microservice systems for software engineering research,'' in {\em Proceedings
  of the 40th International Conference on Software Engineering: Companion
  Proceeedings}, ICSE '18, (New York, NY, USA), pp.~323--324, ACM, 2018.

\bibitem{Ppbench}
N.~Kratzke and P.-C. Quint, ``Ppbench,'' in {\em Proceedings of the 6th
  International Conference on Cloud Computing and Services Science - Volume 1
  and 2}, CLOSER 2016, (Portugal), pp.~223--231, SCITEPRESS - Science and
  Technology Publications, Lda, 2016.

\bibitem{Gan19}
Y.~Gan, Y.~Zhang, D.~Cheng, A.~Shetty, P.~Rathi, N.~Katarki, A.~Bruno, J.~Hu,
  B.~Ritchken, B.~Jackson, K.~Hu, M.~Pancholi, Y.~He, B.~Clancy, C.~Colen,
  F.~Wen, C.~Leung, S.~Wang, L.~Zaruvinsky, M.~Espinosa, R.~Lin, Z.~Liu,
  J.~Padilla, and C.~Delimitrou, ``{An Open-Source Benchmark Suite for
  Microservices and Their Hardware-Software Implications for Cloud and Edge
  Systems},'' in {\em Proceedings of the Twenty Fourth International Conference
  on Architectural Support for Programming Languages and Operating Systems
  (ASPLOS)}, April 2019.

\bibitem{Gan18}
Y.~Gan and C.~Delimitrou, ``{The Architectural Implications of Cloud
  Microservices},'' in {\em Computer Architecture Letters (CAL), vol.17, iss.
  2}, Jul-Dec 2018.

\bibitem{Gan18b}
Y.~Gan, M.~Pancholi, D.~Cheng, S.~Hu, Y.~He, and C.~Delimitrou, ``{Seer:
  Leveraging Big Data to Navigate the Complexity of Cloud Debugging},'' in {\em
  Proceedings of the Tenth USENIX Workshop on Hot Topics in Cloud Computing
  (HotCloud)}, July 2018.

\bibitem{Delimitrou19}
Y.~Gan, Y.~Zhang, K.~Hu, Y.~He, M.~Pancholi, D.~Cheng, and C.~Delimitrou,
  ``{Seer: Leveraging Big Data to Navigate the Complexity of Performance
  Debugging in Cloud Microservices},'' in {\em Proceedings of the Twenty Fourth
  International Conference on Architectural Support for Programming Languages
  and Operating Systems}, April 2019.

\bibitem{thrift}
``Apache thrift.'' \url{https://thrift.apache.org}.

\bibitem{meisner2012bighouse}
D.~Meisner, J.~Wu, and T.~F. Wenisch, ``Bighouse: A simulation infrastructure
  for data center systems,'' in {\em Performance Analysis of Systems and
  Software (ISPASS), 2012 IEEE International Symposium on}, pp.~35--45, IEEE,
  2012.

\bibitem{zeng1998glomosim}
X.~Zeng, R.~Bagrodia, and M.~Gerla, ``Glomosim: a library for parallel
  simulation of large-scale wireless networks,'' in {\em ACM SIGSIM Simulation
  Digest}, vol.~28, pp.~154--161, IEEE Computer Society, 1998.

\bibitem{xie2009aqua}
P.~Xie, Z.~Zhou, Z.~Peng, H.~Yan, T.~Hu, J.-H. Cui, Z.~Shi, Y.~Fei, and
  S.~Zhou, ``Aqua-sim: An ns-2 based simulator for underwater sensor
  networks,'' in {\em OCEANS 2009, MTS/IEEE biloxi-marine technology for our
  future: global and local challenges}, pp.~1--7, IEEE, 2009.

\bibitem{henderson2008network}
T.~R. Henderson, M.~Lacage, G.~F. Riley, C.~Dowell, and J.~Kopena, ``Network
  simulations with the ns-3 simulator,'' {\em SIGCOMM demonstration}, vol.~14,
  no.~14, p.~527, 2008.

\bibitem{issariyakul2012introduction}
T.~Issariyakul and E.~Hossain, ``Introduction to network simulator 2 (ns2),''
  in {\em Introduction to Network Simulator NS2}, pp.~21--40, Springer, 2012.

\bibitem{Harchol13}
M.~Harchol-Balter, {\em Performance Modeling and Design of Computer Systems:
  Queueing Theory in Action}.
\newblock Cambridge University Press, 2013.

\bibitem{Kleinrock}
L.~Kleinrock, ``Queueing systems volume 1: Theory,'' pp. 101-103, 404.

\bibitem{nginx}
``Nginx.'' \url{https://nginx.org/en}.

\bibitem{chen2017workload}
S.~Chen, S.~GalOn, C.~Delimitrou, S.~Manne, and J.~F. Mart{\'\i}nez, ``Workload
  characterization of interactive cloud services on big and small server
  platforms,'' in {\em Workload Characterization (IISWC), 2017 IEEE
  International Symposium on}, pp.~125--134, IEEE, 2017.

\bibitem{Delimitrou15}
C.~Delimitrou, D.~Sanchez, and C.~Kozyrakis, ``{Tarcil: Reconciling Scheduling
  Speed and Quality in Large Shared Clusters},'' in {\em Proceedings of the
  Sixth ACM Symposium on Cloud Computing (SOCC)}, August 2015.

\bibitem{Lo14}
D.~Lo, L.~Cheng, R.~Govindaraju, L.~A. Barroso, and C.~Kozyrakis, ``Towards
  energy proportionality for large-scale latency-critical workloads,'' in {\em
  Proceedings of the 41st Annual International Symposium on Computer
  Architecuture (ISCA)}, Minneapolis, MN, 2014.

\bibitem{Adrenaline}
C.~Hsu, Y.~Zhang, M.~A. Laurenzano, D.~Meisner, T.~F. Wenisch, J.~Mars,
  L.~Tang, and R.~G. Dreslinski, ``Adrenaline: Pinpointing and reining in tail
  queries with quick voltage boosting,'' in {\em 21st {IEEE} International
  Symposium on High Performance Computer Architecture, {HPCA} 2015, Burlingame,
  CA, USA, February 7-11, 2015}, pp.~271--282, 2015.

\bibitem{meisner09}
D.~Meisner, B.~T. Gold, and T.~F. Wenisch, ``Powernap: eliminating server idle
  power,'' in {\em Proceedings of the 14th international ASPLOS}, ASPLOS '09,
  2009.

\end{thebibliography}

\end{document}